\documentstyle[12pt,epsfig]{book}
\makeatletter
\def\nothing#1{}
\newdimen\earraycolsep
\renewcommand{\thetable}{\arabic{table}}

\def\title{\chapter}
\renewcommand\chapter{\ifodd\c@page\clearpage\else\cleardoublepage\fi
		    \global\@topnum\z@
		    \@afterindenttrue
		    \secdef\@chapter\@schapter}
\def\@makechapterhead#1{%
  \vspace*{120\p@}%
  {\parindent \z@ \raggedright \reset@font
    \bfseries #1\par
    \nobreak
    \vskip 36\p@
  }}
\def\author#1
{{\pretolerance=10000 \raggedright \advance 
\leftskip by 1in \noindent #1 \vskip 1pc}}
\def\affiliation#1{{\advance\leftskip by 1in \noindent #1 \vskip -1pc}}

\def\tablenote#1
{\setbox0=\hbox{$^{\hbox{\scriptsize #1}}$}\noindent\hangindent=\wd0 \box0}
\renewcommand\section{\@startsection{section}{1}{\z@}{2pc \@plus 1ex minus
    .2ex}{1pc \@plus .2ex}{\reset@font\normalsize\bfseries}}
\renewcommand\subsection{\@startsection{subsection}{2}{\z@}{1pc \@plus 1ex
    minus.2ex}{1pc \@plus .2ex}{\reset@font\normalsize\bfseries}}
\renewcommand\subsubsection{\@startsection{subsubsection}{3}{\parindent}
	{1pc \@plus 1ex minus.2ex}{-0.5em}{\reset@font\normalsize\bfseries}}

\def\AmS{{\protect\the\textfont2 A\kern-.1667em\lower.5ex\hbox{M}\kern-.125emS}}

\def\p@LaTeX{{\family{times}\series{m}\shape{n}
\selectfont L\kern-.36em\raise.3ex
\hbox{\scriptsize A}\kern-.15em T\kern-.1667em\lower.7ex\hbox{E}\kern-.125emX}}

\newlength{\colwidth}

\def\@oddhead{\hfil}
\def\@evenhead{\hfil}
\def\@oddfoot{{\bfseries\hfil\thepage}}
\def\@evenfoot{{\bfseries\thepage\hfil}}
\def\fnum@table{\normalsize\raggedright{\bfseries \tablename~\thetable.}}
\long\def\@makefntext#1{\setbox0=\hbox{$\m@th^{\@thefnmark}$}\noindent
\hangindent=\wd0 \box0 #1}
\def\centerfig#1#2#3#4{\vspace*{#2}\relax\centerline
{\hbox to#1{\special{#4:#3.#4 x=#1, y=#2}\hfil}}}
\newbox\@atbox
\long\def\atable#1#2#3{\begin{table}[tbp]\centering\footnotesize
\setbox\@atbox\hbox{#2}
\parbox{\wd\@atbox}{\caption{#1}}\par\smallskip #2
\par\smallskip\parbox{\wd\@atbox}{\raggedright #3}
\end{table}}
\def\@bibitem{\noindent \hangindent=2pc \hangafter=1}
\def\thebibliography{%
\section*{REFERENCES}%
\bgroup\footnotesize
\def\newblock{\hskip .11em plus.33em minus.07em}%
\let\bibitem\@bibitem}
\def\endthebibliography{\par\egroup}
\def\@nbibitem#1{\noindent \hangindent=2pc \hangafter=1
\refstepcounter{enumi}\hbox to 2pc{\arabic{enumi}.\hfil}%
\immediate\write\@auxout{\string\bibcite{#1}{\arabic{enumi}}}}
\def\numbibliography{%
\section*{REFERENCES}%
\bgroup\footnotesize
\setcounter{enumi}{0}%
\def\newblock{\hskip .11em plus.33em minus.07em}%
\let\bibitem\@nbibitem}
\def\endnumbibliography{\par\egroup}
\makeatother
\def\beq{\begin{equation}}
\def\eeq{\end{equation}}

\def\beqn{\begin{eqnarray}}
\def\eeqn{\end{eqnarray}}

\makeatletter          
\def\bfseries{\bf}     
\def\@plus{plus}       
\font\scriptsize=cmr7  
\def\p@LaTeX{{\reset@font\rm L\kern-.36em\raise.3ex\hbox{\sc a}\kern-.15em %
     T\kern-.1667em\lower.7ex\hbox{E}\kern-.125emX}}
\makeatother           

\begin{document}

\chapter{SUSY AND SUCH}

\author{S. Dawson}
\affiliation{Physics Department\\
Brookhaven National Laboratory\\
Upton, NY, 11973}

\section{INTRODUCTION}

The Standard Model of particle physics is in stupendous agreement
with experimental measurements; in some cases it has been tested to a 
precision of greater than $.1\%$.  Why then expand our model?  The
reason, of course, is that the Standard Model contains several nagging 
theoretical problems which cannot be solved without the introduction
of some new physics.  Supersymmetry is, at present, many theorists' 
favorite  candidate for such new physics.

The single aspect of the Standard Model which has not been
verified experimentally is the Higgs sector.  
The Standard Model without the Higgs boson is
incomplete, however,
 since  it predicts massless fermions and gauge
bosons.  Furthermore, the electroweak radiative corrections
would be  infinite and longitudinal gauge boson scattering
would  grow
with energy and violate unitarity at an energy scale around
$3~TeV$ if there were no Higgs boson.\cite{lqt}
  The simplest mechanism to cure these defects
is the introduction of a single $SU(2)_L$ doublet of Higgs bosons.
When the neutral component of the Higgs boson gets a 
vacuum expectation value, the $SU(2)_L\times U(1)_Y$ gauge
symmetry is broken, giving the $W$ and $Z$ gauge bosons
their masses.  The chiral symmetry forbidding fermion masses
is broken at the same time allowing the fermions to become
massive. Furthermore, in the Standard Model, the coupling of the Higgs
boson to gauge bosons is just that required to cancel
the  infinities
in electroweak radiative corrections and to cancel the unitarity
violation in the gauge boson scattering sector.  What then is the
problem with this simple and economical picture?

The argument against the simplest  version of the Standard Model
with a single Higgs boson is purely theoretical and arises
when radiative corrections to the Higgs boson mass are computed.
The scalar potential  for the Higgs boson, $h$,
 is given schematically by,
\beq
 V\sim M_{h0}^2 h^2 + \lambda h^4. 
\eeq 
 At one loop, the quartic self- interactions
of the Higgs boson (proportional to $\lambda$)
 generate a quadratically
divergent contribution
 to the Higgs boson mass which must be cancelled
by the mass  counterterm, $\delta M_h^2$,\cite{thoof} 
\beq
M_h^2\sim M_{h0}^2 +{\lambda\over 4 \pi^2}\Lambda^2 
+\delta M_h^2 .   
\eeq
The scale $\Lambda$ is a cutoff which, in the Standard Model
with no new physics between the electroweak scale and the
Planck scale, must be of order the Planck scale.  
In order for the Higgs boson to do its job of preventing
unitarity violation in the scattering of longitudinal gauge
bosons, however,
its mass must be less than around $800~GeV$.\cite{lqt}  
This leads to an unsatisfactory situation.  The large quadratic 
contribution to the Higgs boson mass-squared,
of ${\cal O}(10^{18} GeV)^2$,
 must be cancelled by the
counterterm $\delta M_h^2$ such that the result is 
roughly  less than
$(800~GeV)^2$.
  This requires a cancellation of one  part in
$10^{16}$.  This is of course formally possible, but regarded
by most theorists as an unacceptable fine tuning
of parameters. Additionally,
this cancellation must occur at every order in perturbation
theory and so the parameters
must be fine tuned again and again.    The quadratic growth
of the Higgs  boson mass beyond  tree level in
perturbation theory  is one of
 the driving motivations  behind
the introduction of supersymmetry, which we will see cures this
problem.  
It is interesting that the loop corrections to fermion masses
do not exhibit this quadratic growth (and we therefore say that
fermion masses are ``natural'').  It is only when we attempt
to understand electroweak symmetry breaking by including a
Higgs boson that we face the problem of quadratic divergences.  

In these lectures, I discuss the theoretical motivation for supersymmetric
theories and introduce the minimal low energy effective supersymmetric
theory, (MSSM).
I consider only the MSSM and its simplest grand unified extension
here.  Some of the other possible low-energy SUSY models
are summarized in Ref. \cite{snow}.  
  The particles and their interactions are examined in detail 
in the next sections and
a  grand unified SUSY model presented which
gives additional motivation for pursuing supersymmetric theories.
   
Finally, I discuss indirect limits on the SUSY partners
of ordinary matter coming
from precision measurements at LEP and 
direct production searches at the Tevatron
 and  discuss search strategies for SUSY
at both future  $e^+e^-$ and hadron colliders.
Only  a sampling of existing limits are given in order  to
demonstrate some of the general features of these
searches.  Up- to- date limits on SUSY particle
searches at hadron colliders\cite{mer} and
$e^+e^-$ colliders\cite{sch} were given at the 1996 DPF meeting
and can be used to map out the allowed regions
of SUSY parameter space.    There exist numerous excellent
reviews of both the more formal aspects of supersymmetric model
building \cite{hkrep, bagtasi}
 and the phenomenology of these
models \cite{xerxes, peskin} and the reader is referred to
these for more details.  I present here a workmanlike approach
designed primarily
for experimental graduate students.

\section{WHAT IS SUSY? }

Suppose we reconsider the one loop contributions to the Higgs
boson mass in a theory which contains both massive  scalars,
$\phi$,  and fermions, $\psi$,
 in addition to the Higgs field, $h$.  Then the Lagrangian is given
by:
\beq
{\cal L}\sim -g_F {\overline \psi} \psi h
-g_S^2 h^2 \phi^2
\quad . 
\eeq
If we again calculate the  one-loop
contribution to $M_h^2$ we find\cite{thoof}
\beqn
M_h^2 &\sim&
 M_{h0}^2 +{g_F^2\over 4 \pi^2}\biggl(
\Lambda^2 +m_F^2\biggr)
-{g_S^2\over 4 \pi^2} \biggl( \Lambda^2+m_S^2\biggr)
\nonumber \\
&&\quad\quad +{\rm logarithmic~ divergences} +
 {\rm  uninteresting~ terms } \quad .
\label{mquad} 
\eeqn
The relative minus sign between the fermion and scalar contributions
 to the Higgs boson mass-squared 
is the well-known result of Fermi statistics.  
We see that if $g_S=g_F$ the terms which grow with $\Lambda^2$
cancel and we are left with a well behaved contribution to
the Higgs boson mass so long as the fermion and scalar masses
are not too different,\cite{ander}
\beq
M_h^2\sim M_{h0}^2 +{g_F^2\over 4 \pi^2}\biggl(
m_F^2-m_S^2\biggr)
 \quad .
\eeq
Attempts have been made to quantify ``{\it not too
different}''.[9] 
 One can roughly assume that the cancellation
is unnatural if the mass splitting  between the
fermion and the scalar is larger than 
about a TeV.   Of course, in order for this cancellation
to persist to all orders in perturbation
theory it must be the result of a symmetry.
This symmetry is ${\it{\bf supersymmetry}}$.

Supersymmetry is a symmetry which relates particles of differing
spin, (in the above example,
fermions and scalars).
  The particles are combined into a ${\it superfield}$,
which contains fields differing by one-half unit of spin.\cite{wess}
  The
simplest example, the scalar superfield, contains a complex
scalar, $S$, and a two- component Majorana fermion, $\psi$.
(A Majorana fermion, $\psi$, is one which
is equal to its  charge conjugate,
$\psi^c=\psi$. A familiar example is a  Majorana neutrino.)
The supersymmetry completely specifies the allowed interactions.
In this simple case, the Lagrangian is
\beqn 
{\cal L}&=& -\partial_\mu S^*\partial^\mu S - i
{\overline \psi}{\overline \sigma}^\mu \partial_\mu\psi
-{1\over 2} m(\psi\psi+{\overline \psi}
{\overline\psi})
\nonumber \\
&&-c S\psi\psi-c^* S^*{\overline \psi}{\overline \psi}
-\mid m S+c S^2\mid ^2
 ,  
\eeqn 
(where $\sigma$ is a $2\times 2$ Pauli matrix and $c$
is an arbitrary coupling constant.)
This Lagrangian is invariant (up to a total derivative)
under transformations which take the scalar into the
fermion and ${\it vice~versa}$. 
Since  the scalar and fermion interactions have the
same coupling, the cancellation of quadratic divergences
occurs automatically, as in Eq. \ref{mquad}.  
One thing that is immediately obvious is that this Lagrangian
contains both a scalar and a fermion ${\it of~equal ~mass}$.
Supersymmetry connects particles of different spin, but
with all other
characteristics the same.  That is, they have
the same quantum numbers and the same mass.
\begin{itemize}
\item
Particles in a superfield have the same masses and
quantum numbers and differ by $1/2$ unit of spin in 
a theory with unbroken supersymmetry.
\end{itemize}

It is clear, then, that {\bf
supersymmetry must be a broken symmetry}.
There is no scalar particle, for example, with the mass and
quantum numbers of the electron.  In fact, there are no candidate
supersymmetric scalar partners for any of the fermions in the
experimentally observed 
spectrum.  We will take a non-zero
 mass splitting between the particles of
a superfield as a signal for supersymmetry breaking.

Supersymmetric theories are easily constructed according to the rules
of  supersymmetry.  I present here a cookbook approach to
constructing the minimal supersymmetric version
of the Standard Model.  The first step is to pick the particles in
superfields.  There are two types of superfields relevant for
our purposes:\footnote{
The superfields also contain ``auxiliary fields'', which
are fields with no kinetic energy terms in
the Lagrangian.\cite{wess}  These fields are not important
for our purposes.} 
\begin{enumerate}
\item  ${\it Chiral ~Superfields}$:  These consist of 
a complex scalar field, $S$, and a $2$-component Majorana fermion
field, ${\psi}$.
\item ${\it Massless Vector~Superfields}$:  These consist of a massless
gauge field with field strength $F_{\mu\nu}^A$
 and a $2$-component Majorana fermion field, $\lambda_A$, termed
a ${\it gaugino} $. 
The index $A$ is the gauge index.  
\end{enumerate}

\subsection{The Particles of the MSSM}

The MSSM respects the same $SU(3)\times
SU(2)_L\times U(1)$ gauge symmetries as does the
Standard Model.  
The particles necessary to construct the supersymmetric 
version of the Standard Model are shown in Tables 1 and 2
in terms of the superfields, (which are denoted by the
superscript ``hat'').  
Since there are no candidates for supersymmetric partners of
the observed particles, we must double the entire spectrum,
placing the observed particles in superfields with new
postulated superpartners.  
 There are, of course, quark and
lepton superfields for all $3$ generations and we have
listed  in Table 1
only the members of the first generation.  
The superfield ${\hat Q}$ thus consists of an $SU(2)_L$
doublet of quarks:
\beq
Q= 
\left( \begin{array}{c} u \\
d\end{array}\right)_L
\eeq
and their scalar partners which are also in
an $SU(2)_L$ doublet,
\beq
{\tilde Q}=
\left(\begin{array}{c}  {\tilde u}_L\\ 
{\tilde d}_L\end{array}\right)
\quad .  
\eeq   Similarly, the
superfield ${\hat U}^c$ (${\hat D}^c$)
 contains the right-handed
up  (down) anti-quark, ${\overline u}_R$ (${\overline d}_R$), 
 and its scalar partner, ${\tilde u}_R^*$ (${\tilde d}_R^*$).
The scalar partners of the quarks are fancifully called
squarks.  
We see that each quark has $2$ scalar partners, one corresponding
to each quark chirality.
The leptons are contained in the $SU(2)_L$ doublet superfield
${\hat L}$ which contains the left-handed fermions,
\beq
L=\left(\begin{array}{c} \nu \\
e\end{array}\right)_L
\eeq
and their scalar partners,
\beq
{\tilde L}=\left(\begin{array}{c}
{\tilde \nu}_L\\
{\tilde e}_L\end{array}\right)
\quad .
\eeq
Finally, the right-handed anti-electron, ${\overline e}_R$, is contained
in the superfield ${\hat E}^c$ and has a scalar partner
${\tilde e}_R^*$.  The scalar partners of the leptons
are termed sleptons.  

  The $SU(3)\times SU(2)_L\times U(1)$  gauge fields all obtain
Majorana fermion partners in a SUSY model.
  The ${\hat G}^a$ superfield contains
the gluons, $g^a$, and their partners the gluinos, ${\tilde g}^a$;
${\hat W}_i$ contains the $SU(2)_L$ gauge bosons, $W_i$ and
their fermion partners, ${\tilde \omega}_i$ (winos);
and ${\hat B}$ contains the $U(1)$ gauge field, $B$,
and its fermion partner, ${\tilde b}$ (bino).  The usual
notation is to denote the supersymmetric partner of a fermion
or gauge field with the same letter, but with a tilde over it.   
\begin{table}[htb]
\begin{center}
{Table 1: Chiral Superfields of the MSSM}
\vskip6pt
\renewcommand\arraystretch{1.2}
\begin{tabular}{|lccrc|}
\hline
\multicolumn{1}{|c}{Superfield}& SU(3)& $SU(2)_L$& $U(1)_Y$
& Particle Content 
\\
\hline
${\hat Q}$   &    $3$          & $2$&  $~{1\over 6}$
& ($u_L,d_L$), (${\tilde u}_L,{\tilde d}_L$)\\
${\hat U}^c$ & ${\overline 3}$ & $1$& $-{2\over 3}$
&${\overline u}_R$, ${\tilde u}_R^*$\\
${\hat D}^c$ & ${\overline 3}$ & $1$&  $~{1\over 3}$
&${\overline d}_R$, ${\tilde d}_R^*$\\
${\hat L}$   & $1$             & $2$& $~-{1\over 2}$
& $(\nu_L,e_L)$, (${\tilde \nu}_L, {\tilde e}_L$)\\
${\hat E}^c$ & $1$             & $1$& $~1$ 
& ${\overline e}_R$, ${\tilde e}_R^*$\\
${\hat H_1}$ & $1$             & $2$& $-{1\over 2}$ 
&($H_1, {\tilde h}_1$)\\
${\hat H_2}$ & $1$             & $2$& $~{1\over 2}$
& $(H_2, {\tilde h}_2)$ \\ 
\hline
\end{tabular}
\end{center}
\end{table}
 
\begin{table}[htb]
\begin{center}
{Table 2: Vector Superfields of the MSSM}
\vskip6pt
\renewcommand\arraystretch{1.2}
\begin{tabular}{|lcccc|}
\hline
\multicolumn{1}{|c}{Superfield}&SU(3)&$SU(2)_L$&$U(1)_Y$
& Particle Content\\
\hline
${\hat G^a}$  &  $8$  &  $1$  &  $0$ 
&$g$, ${\tilde g}$ \\
${\hat W^i}$  &  $1$  &  $3$  &  $0$ 
& $W_i$, ${\tilde \omega}_i$  \\
${\hat B}$  &  $1$  &  $1$  &  $0$ 
& $B$, ${\tilde b}$  \\
\hline
\end{tabular}
\end{center}
\end{table}

One feature of Table 1 requires explanation.  The Standard Model
contains a single $SU(2)_L$ doublet of scalar particles, dubbed the
``Higgs doublet".  In the supersymmetric extension of
the Standard Model, this scalar doublet acquires a SUSY
partner which is an $SU(2)_L$ doublet of
Majorana  fermion fields, 
 ${\tilde h}_1$  (the Higgsinos), which   
 contribute to the triangle $SU(2)_L$ and $U(1)$ gauge anomalies.
Since the fermions of the Standard Model have exactly the
right quantum numbers to cancel these anomalies, it follows
that the contribution from the fermionic partner of the
Higgs doublet remains uncancelled.\cite{anoms}    
Since gauge theories cannot have anomalies, these contributions
must be cancelled somehow if the SUSY theory is to be sensible.
The simplest way is to add a second Higgs doublet with 
precisely the opposite $U(1)$ quantum numbers from the
first Higgs doublet.  In a SUSY Model,
 this second
Higgs doublet will also have fermionic partners, ${\tilde h}_2$,
 and the contributions of the fermion partners of the
two Higgs doublets to gauge anomalies
will precisely cancel each other, leaving
an anomaly free theory.  
It is easy to check that the fermions of Table 1 satisfy the
conditions for anomaly cancellation:
\beq
Tr(Y^3)=Tr(T_{3L}^2Y)=0\quad .
\eeq 
We will see later that $2$ Higgs doublets are also required in order to 
give both the up and down quarks masses in a SUSY theory.  
The requirement that there be at least $2$  $SU(2)_L$
Higgs doublets is a feature
of all models with weak scale supersymmetry.  
\begin{itemize}
\item  
In  general,  supersymmetric extensions of the
Standard Model  have extended Higgs sectors leading to a rich
phenomenology of scalars.  
\end{itemize}

\subsection{The Interactions of the MSSM} 
Having specified the superfields of the theory, the next
step is to construct the supersymmetric Lagrangian.\cite{early}  There
is very little freedom in the allowed interactions between
the ordinary particles and their supersymmetric partners.
It is this feature of a SUSY model which gives it predictive
power (and makes it attractive to theorists!). 
It is important to note here, however,
 that there is nothing to stop us from
adding more superfields to those shown in Tables 1 and 2 as
long as we are careful to add them in such a way that 
any  additional contributions to gauge anomalies cancel
among themselves.  Recent
popular models add an additional gauge singlet superfield
to the
spectrum, which has interesting phenomenological consequences.\cite{mess}  
The MSSM which we concentrate on,
 however, contains only those fields given in the tables.    

The supersymmetry  associates  each $2$-component Majorana fermion
with a complex scalar.  The massive
fermions of the Standard Model are, however,
Dirac fermions.   A Dirac fermion has $4$
components which can be thought of as the left-and right-handed
chiral projections of the fermion state.
  It is straightforward to translate back
and forth between $2$- and $4$- component notation for the fermions
and we will henceforth use the more familiar $4$- component
notation when writing the fermion interactions.\cite{hkrep}   
The fields of the MSSM all have canonical kinetic energies:\footnote{
Remember that both the right- and left- handed helicity state of
a fermion has its own scalar partner.} 
\begin{eqnarray} 
{\cal L}_{KE}&=&\sum_i\biggl\{ (D_\mu S_i^*)(D^\mu S_i)
+i{\overline \psi}_i D \psi_i\biggr\}\nonumber \\
&&+\sum_A\biggl\{ -{1\over 4} F_{\mu\nu}^A F^{\mu\nu A}
+{i\over 2} {\overline \lambda_A} D \lambda_A
\biggr\}
, 
\eeqn 
where $D$ is the $SU(3)\times SU(2)_L\times U(1) $
gauge invariant derivative.  
  The $\sum_i$ is over all the fermion fields of the Standard
Model, $\psi_i$, and their scalar partners, $S_i$,    
 and also over the $2$ Higgs
doublets with their fermion partners.  The $\sum_A$  is over
the $SU(3)$, $SU(2)_L$ and $U(1)_Y$ gauge fields with their
fermion partners, the gauginos.  
  
The interactions between the chiral superfields of Table 1 
  and the
gauginos  and the gauge fields
 of Table 2
are completely specified by the gauge symmetries
and by the supersymmetry, as are the quartic interactions of
the scalars,                
\beq
{\cal L}_{int}=-\sqrt{2}\sum_{i,A}
 g_A\biggl[S_i^* T^A {\overline \psi}_{iL}
\lambda_A +{\rm h.c.}\biggr] -{1\over 2} 
\sum_A \biggl( \sum_i g_A S_i^* T^A S_i\biggr)^2
\quad ,  
\label{scalints}  
\eeq       
where $\psi_L\equiv {1\over 2} (1-\gamma_5)\psi $.  
In  Eq. \ref{scalints}, $g_A$ is the relevant gauge
coupling constant and we see that the interaction strengths
are fixed in terms of these constants. 
{\bf There are no adjustable parameters here.}
 For example, the
interaction between a quark, its scalar partner, the squark,  
 and the gluino is governed
 by the strong coupling
constant, $g_s$. A  complete set of Feynman rules for the minimal
SUSY model described here is given in the review by Haber
and Kane.\cite{hkrep}  A good rule of thumb is to take an interaction
involving Standard Model particles
 and replace two of the particles
by their SUSY partners to get an approximate strength for
the  interaction.  (This naive picture is, of course, altered by 
$\sqrt{2}$'s, mixing angles, etc.).

The only freedom in constructing the supersymmetric Lagrangian
(once the superfields and the gauge symmetries are chosen)
is contained in a function called the ${\it {\bf superpotential}}$,$W$.
The superpotential is a function of the chiral superfields
of Table 1 only
(it is not allowed to contain their complex congugates) and it
contains terms with $2$ and $3$ chiral superfields.  Terms in
the superpotential with more than $3$ chiral superfields would
yield non-renormalizable interactions in the Lagrangian.
The superpotential also
 is not allowed to contain derivative interactions and we
say that it is an analytic function.
  From
the superpotential can be found both the scalar potential and
the Yukawa interactions of the fermions with the scalars:
\beq 
{\cal L}_{W}=-\sum_i \mid {\partial W\over
\partial z_i}\mid ^2 -{1\over 2}\sum_{ij}
\biggl[ {\overline \psi}_{iL} {\partial^2 W
\over \partial z_i \partial z_j}\psi_j+{\rm
h.c.}\biggr]
, 
\label{lagw} 
\eeq
where $z$ is a chiral superfield.
This form of the Lagrangian is dictated by the supersymmetry
and by the requirement that it be renormalizable.  An explicit
derivation of Eq. \ref{lagw} can be found in Ref. \cite{wess}.
  To obtain the interactions, we take the derivatives of $W$ with
respect to the superfields, $z$, and then evaluate the result in 
terms of
the scalar component of $z$.   
  
The usual approach is to write the most general
$SU(3)\times SU(2)_L\times U(1)_Y$ invariant 
superpotential with arbitrary coefficients for the interactions,  
\begin{eqnarray}
W &=& \epsilon_{ij} \mu {\hat H}_1 ^i {\hat H}_2^j 
+\epsilon_{ij}
\biggl[ \lambda_L  {\hat H}_{1 }^i {\hat L}^{cj}{\hat E}^c +
\lambda_D  {\hat H}_1^i {\hat Q}^j {\hat D}^c
+\lambda_U  {\hat H}_2^j {\hat Q}^i {\hat U}^c\biggr]
\nonumber \\
&& + \epsilon_{ij}\biggl[ {\lambda_1} {\hat L}^i 
{\hat L}^j {\hat E}^c +
\lambda_2 {\hat L}^i {\hat Q}^j {\hat D}^c
\biggr] 
+\lambda_3 {\hat U}^c {\hat D}^c {\hat D}^c
, 
\label{superpot} 
\eeqn  
(where  $i,j$ are $SU(2)$ indices).
In principle, a bi-linear term $\epsilon_{ij}
{\hat L}^i{\hat H_2}^j$ can also be included in the
superpotential.  It is possible, however, to rotate the
lepton field, ${\hat L}$, such that this term vanishes so we
will ignore it.   
We have written the superpotential in terms of the fields of the first
generation.  In principle, the $\lambda_i$
 could all be  matrices which
mix the interactions of the $3$ generations.

The $\mu {\hat H}_1 {\hat H}_2$ term in the superpotential
gives mass terms for the
Higgs bosons when we apply $\mid \partial W/\partial z\mid^2$
 and $\mu$ is often called the Higgs mass
parameter.   
We shall see later that the physics
is very sensitive to the sign of $\mu$.  The
terms in the square brackets proportional
to $\lambda_L$, $\lambda_D$, and $\lambda_U$  give the usual
Yukawa interactions of the fermions with the
Higgs bosons from the term
${\overline \psi}_i (\partial^2 W/\partial z_i 
\partial z_j) \psi_j$.
 Hence   these coefficients are
determined  in terms of the fermion masses and
the 
vacuum expectation values of the neutral
members of the scalar components of the Higgs doublets and are not
free parameters at all.

The Lagrangian as we have written it cannot, however, be the
whole story as all the particles (fermions, scalars,
gauge fields) are massless at this point.
 
\subsection{R Parity}  
The terms in the second line of Eq. \ref{superpot} (proportional
to $\lambda_1, \lambda_2$ and $\lambda_3$) are
a problem.  They contribute to  lepton and
baryon number violating interactions and
can mediate proton decay at tree level through the exchange
of the scalar partner of the down quark. 
If the SUSY partners of the Standard Model 
particles have masses on the TeV scale, then
these interactions are severely restricted by 
experimental measurements.\cite{early,proton}

There are several possible approaches to the
problem of the lepton and baryon number violating interactions.
The first is simply to make the coefficients, $\lambda_1,
\lambda_2$, and $\lambda_3$ 
small enough to avoid experimental limits.\cite{sher,rparity}   This artificial
tuning of parameters is regarded as unacceptable  by 
many theorists, but is certainly
allowed experimentally.  Another tactic is to
make either  the lepton number violating
interactions, $\lambda_1$ and $\lambda_2$, or
the baryon number violating interaction,  $\lambda_3$, zero,
(while allowing the others to be non-zero) which
would  forbid proton decay.  
There is, however, not much theoretical motivation for this
approach.

  The usual strategy   is to require
that  all of these undesirable  lepton and baryon number
violating terms be forbidden by a symmetry.
(If they are forbidden by a symmetry, they will not
re-appear at higher orders of perturbation theory.)  
The symmetry which does the job is 
called ${\it\bf R~parity}$.\cite{rp} 
R parity can be defined as a multiplicative quantum number
such that all particles of the Standard Model have R parity
+1, while their SUSY partners have R parity -1.
R parity can also be defined as,
\beq
R\equiv (-1)^{3(B-L)+s}
\quad ,
\eeq
for a particle of spin $s$.     It is
then obvious  that such a symmetry forbids the lepton
and baryon number violating terms of Eq. \ref{superpot}. 
 It is worth
noting that in the Standard Model, the problem
of baryon and lepton number violating interactions  does not
arise, since these  interactions are 
forbidden by the gauge symmetries  to contribute
to dimension- $4$ operators
and first arise in dimension- $6$ operators which are
suppressed by factors of some heavy mass scale.

The assumption of R parity conservation has profound
experimental consequences which go beyond the details
of a specific model.  Because R parity is a multiplicative
quantum number, it implies that the number of SUSY partners
in a given interaction is always conserved modulo 2.  
\begin{itemize}
\item
 SUSY
partners can only be pair produced from  Standard Model
particles.  
\end{itemize}
Furthermore, a SUSY particle will decay in a chain until
the lightest SUSY particle is produced (such a decay is
called a ${\it cascade~decay}$).  This lightest SUSY
particle, called the LSP, must be absolutely stable when R
parity is conserved.
\begin{itemize}
\item
A theory with $R$ parity conservation will have a 
lightest SUSY particle (LSP) which is stable.  
\end{itemize}
 The LSP must be neutral since there
are stringent cosmological bounds on  light
charged or colored
particles which are stable.\cite{pdg,lsplims}  Hence the LSP is
stable and  neutral and
is not seen in a detector (much like a neutrino)
since it interacts only by the exchange of a heavy virtual SUSY
particle.  
\begin{itemize}
\item  The LSP will interact very weakly with
	ordinary  matter.  
\item  
A generic signal for R parity conserving SUSY theories 
is missing transverse energy from the non-observed LSP.  
\end{itemize}  
In theories without $R$ parity conservation, there will
not be a stable  LSP, and the lightest SUSY
particle will decay into ordinary particles (possibly
within the detector).   Missing transverse energy 
will no 
longer be a robust signature for SUSY particle production.\cite{baer1}

\subsection{Supersymmetry Breaking}  

The mechanism of supersymmetry breaking is not well understood.
At this point we have 
constructed a SUSY theory containing all of the
Standard Model particles, but the supersymmetry
remains unbroken and the particles and their SUSY partners 
are massless. This is clearly unacceptable.  
It is typically
assumed that the SUSY breaking occurs
 at a high scale, say $M_{pl}$, and perhaps
results from some complete theory encompassing gravity.  At the
moment the usual approach is  to assume that the MSSM, which
is the theory at the electroweak scale, is an effective
low energy theory.\cite{wein} 
The supersymmetry breaking is implemented by including  explicit
``soft''
 mass terms for the scalar members of the
chiral multiplets and for the gaugino members of the vector
supermultiplets in the
Lagrangian.
  These interactions are termed soft because
they do not re-introduce the quadratic divergences which motivated
the introduction of the supersymmetry in the first place.  
The dimension of soft operators in the Lagrangian must be
$3$ or less, which means that the possible
 soft operators are mass terms,
bi-linear mixing terms (``B'' terms), and 
tri-linear scalar mixing terms (`` A terms'').  
The origin of these supersymmetry breaking terms is left
unspecified.  
The complete set of soft SUSY breaking terms
(which respect R parity and the $SU(3)\times SU(2)_L\times U(1)$
gauge symmetry)
for the first generation  
 is given by the Lagrangian:\cite{early,soft} 
\beqn
-{\cal L}_{soft}&=&
m_1^2 \mid H_1\mid^2 +m_2^2 \mid H_2\mid^2 - B \mu 
\epsilon_{ij} (H_1^i H_2^j + {\rm h.c.}) 
+{\tilde M_Q}^2 ({\tilde u_L}^* {\tilde u_L}
+{\tilde d_L}^* {\tilde d_L})
\nonumber \\  && +{\tilde M_u}^2 
{\tilde u_R}^* {\tilde u_R} +{\tilde M_d}^2
{\tilde d_R}^* {\tilde d_R} 
+
{\tilde M_L}^2({\tilde e_L}^*{\tilde e_L}
+{\tilde \nu_L}^*{\tilde \nu}_L)
+{\tilde M_e}^2 {\tilde e}_R^*{\tilde e_R}
\nonumber \\  &&
+{1\over 2}\biggl[ M_3 {\overline {\tilde g}} {\tilde g}
+M_2 {\overline {\tilde \omega_i}}{\tilde \omega_i}
+M_1{\overline {\tilde b}}{\tilde b}
\biggr] 
+{g\over \sqrt{2}M_W}\epsilon_{ij}
\biggl[ {M_d\over \cos\beta}A_dH_1^i
{\tilde Q}^j {\tilde d_R}^*
\nonumber \\  && +
{M_u\over \sin\beta}A_u H_2^j {\tilde Q}^i
{\tilde u_R}^*   
+{M_e\over \cos\beta}A_e H_1^i{\tilde L}^j {\tilde e}_R^*
+{\rm h.c.} \biggr]
\quad .
\label{lagsoft} 
\eeqn
This Lagrangian has arbitrary masses for the scalars and
gauginos and also arbitrary tri-linear and bi-linear
mixing terms.  
The scalar and gaugino mass terms 
have the desired effect of breaking the degeneracy 
between the particles and their SUSY partners.
The tri-linear A-terms have been defined with an
explicit factor of mass and we will see later that they
affect primarily the particles of the third generation.\footnote{
We have also included an angle $\beta$ in the normalization of the $A$
terms.  The factor
$\beta$ is related to the vacuum expectation values
of the neutral components of the Higgs fields and is defined in
the next section.  The normalization is, of course, arbitrary.}    
When the $A_i$  terms are non-zero, the scalar partners of the
left- and right-handed fermions can mix
when the Higgs bosons get vacuum
expectation values  and so they  are no longer mass
eigenstates.  The $B$ term mixes the scalar components of the $2$  
Higgs doublets.   

The philosophy is to add all of the mass and mixing terms
which are allowed by the gauge symmetries.  
To further complicate matters, all of the mass  and interaction terms 
 of Eq. \ref{lagsoft} 
 may be  matrices involving all three generations.  
${\cal L}_{soft}$ has clearly broken the supersymmetry
since the SUSY partners of the ordinary particles have
been given arbitrary masses.  This has come at the
tremendous
expense, however, of introducing a large number of
unknown parameters (more than 50!). 
 It is one of the wonderful features
of supersymmetry that even with all these new parameters,
the theory is still able to make some definitive predictions.  
This is, of course, because the gauge interactions of the
SUSY particles are completely fixed.  
What is really needed, however, is a theory of how the soft SUSY 
breaking terms arise in order to reduce the parameter
space.  

We have now constructed the Lagrangian describing
a softly broken supersymmetric theory which is
assumed to be the effective theory at the weak scale.
A more complete theory would predict the soft SUSY 
breaking terms.  In the next
section we will examine how the electroweak symmetry
is broken in this model and study the mass spectrum and
interactions of the new particles.  

\subsection{The Higgs Sector and Electroweak Symmetry Breaking}

The Higgs sector of the MSSM is very similar to that of
a general $2$ Higgs doublet model.\cite{hks}  The scalar potential
involving the Higgs bosons  is
\beqn
V_H&=&
\biggl(\mid \mu\mid^2 +m_1^2\biggr)\mid H_1\mid^2
+\biggl(\mid \mu \mid^2+m_2^2\biggr)\mid H_2\mid^2
-\mu B \epsilon_{ij}\biggl(H_1^i H_2^j+{\rm h.c.} \biggr)
\nonumber \\
&&
+{g^2+g^{\prime 2}\over 8}\biggl(
\mid H_1\mid^2 - \mid H_2\mid^2\biggr)^2
+{1\over 2} g^2 \mid H_1^*H_2\mid^2
\quad .
\label{higgspot} 
\eeqn 
The Higgs potential of the SUSY model can be seen
to depend on $3$ independent parameters, 
\beqn 
&& \mid\mu\mid^2+m_1^2,
\nonumber \\
&& 
\mid \mu \mid^2+m_2^2,
\nonumber \\
&&~~~  \mu B~~, 
\eeqn   
where $B$ is a new mass parameter.  
This is in 
contrast to the general $2$ Higgs doublet model where 
there are $6$ arbitrary coupling constants (and a phase)
 in the potential.  
From Eq. \ref{scalints}, it is clear that the quartic couplings
are fixed in terms of the gauge couplings and so they are
not free parameters.  
This leaves only the mass terms of Eq. 20 unspecified. 
Note that $V_H$ automatically conserves CP since
any complex phase in $\mu B$ can be absorbed into the
definitions of the Higgs fields.    
  
Clearly, if $\mu B=0$ then all the terms in the potential
are positive and the minimum  of the potential
occurs with $V=0$ and $\langle H_1^0\rangle=\langle H_2^0
\rangle=0$,  leaving
the electroweak symmetry unbroken.\footnote{
It also leaves the supersymmetry unbroken,
since $\langle V\rangle > 0$ is required
in order for the supersymmetry to be broken.[25]}
  Hence all $3$ parameters
must be non-zero in order for the electroweak symmetry to be
broken.  
\footnote{
We assume that the parameters are arranged in such
a way that the scalar partners of the quarks and leptons
do not obtain vacuum expectation values.  Such vacuum
expectation values would spontaneously break the $SU(3)$ 
color  gauge symmetry or lepton number.
This requirement gives a restriction on
\protect $ A_i/{\tilde m}$, where 
\protect 
${\tilde m}$ is a generic squark or slepton mass.}
                                                      
In order for the electroweak symmetry to be broken
and for the potential to be stable at large values
of the fields, the
parameters must satisfy the relations,
\beqn
(\mu B)^2 & >&\biggl(\mid \mu\mid^2+m_1^2\biggr)
\biggl(\mid \mu\mid^2+m_2^2\biggr)
\nonumber \\
\mid \mu\mid^2+{m_1^2+m_2^2\over 2}& >& \mid \mu B\mid
\quad .
\eeqn
We will assume that these conditions are  met.  
The symmetry is broken when the neutral components of the Higgs doublets
get vacuum expectation values,\footnote{Our  conventions for factors
of $2$ in the Higgs sector, and for the definition of the 
sign$(\mu)$,  are those of Ref. \cite{hhg}.} 
\beqn
\langle H_1^0\rangle & \equiv & v_1
\nonumber \\
\langle H_2^0\rangle & \equiv & v_2
\quad . 
\eeqn  
By redefining the Higgs fields, we can always 
choose $v_1$ and $v_2$ positive.

\begin{figure}[htb]
\vspace*{1in}  
\centerline{\epsfig{file=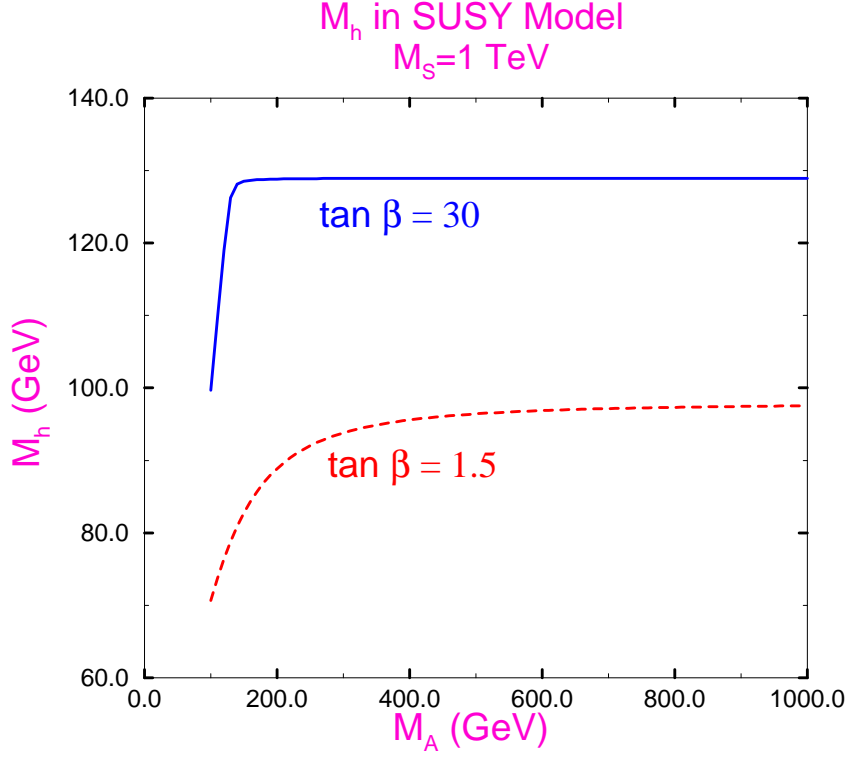,height=4.in}}
\caption{Mass of the lightest neutral Higgs boson as
a function of the pseudoscalar mass, $M_A$,
and $\tan\beta$. This figure includes radiative corrections
to the Higgs mass[28],  assumes a common scalar mass of
$1~TeV$, and neglects mixing effects, ($A_i=\mu=0$).}
\vspace*{.5in}  
\end{figure} 

When the electroweak symmetry is broken,
the $W$ gauge boson gets a mass which
is fixed by $v_1$ and $v_2$,
\beq
M_W^2={g^2\over 2}(v_1^2+v_2^2)\quad .\eeq
Before the symmetry was broken, the $2$ complex $SU(2)_L$ 
Higgs doublets had $8$ degrees of freedom.  Three of
these were  absorbed to give the $W$ and $Z$ gauge bosons
their masses, leaving $5$ physical degrees of freedom.
There is now a charged Higgs boson, $H^\pm$, a CP -odd neutral
Higgs boson, $A$, and $2$ CP-even neutral Higgs bosons, $h$ and $H$.
After fixing $v_1^2+v_2^2$ such that the $W$ gets the correct
mass, the Higgs sector is  then described by $2$ additional
parameters which can be chosen however you like.  The
usual choice is
\beq
\tan\beta\equiv {v_2\over v_1}\eeq
and $M_A$, the mass of the pseudoscalar Higgs boson.
Once these two parameters are given, then the  masses of
the remaining Higgs bosons can be calculated in terms
of $M_A$ and $\tan\beta$.
Note that we can chose $0 \le \beta\le {\pi\over 2}$ since we have
chosen $v_1, v_2 > 0$.

\begin{figure}[htb]
\vspace*{1.in} 
\centerline{\epsfig{file=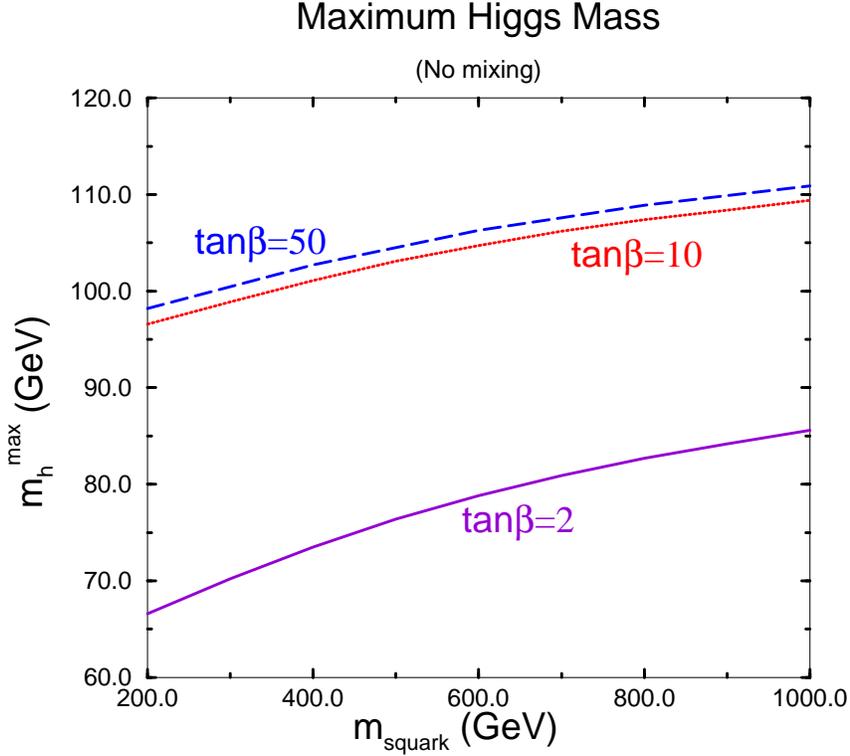,height=4.in}}
\caption{Maximum value of the lightest Higgs boson  mass as a
function of the  squark mass including radiative corrections.[28]
(We have assumed degenerate squarks and set  the
mixing parameters $A_i=\mu=0$.)}  
\vspace*{.5in}
\end{figure}

It is straightforward to find the physical Higgs bosons 
and their masses in terms of the parameters of 
Eq. \ref{higgspot}.  Details can be found in Ref. \cite{hhg}.  
The neutral Higgs masses are found by diagonalizing
the $2\times 2$ Higgs mass matrix and    
by convention, $h$ is taken to be the lighter of the neutral
Higgs.
The pseudoscalar mass is given by,
\beq 
M_A^2={2 \mid \mu B \mid \over \sin 2 \beta},
\eeq
and the charged scalar mass is,
\beq
M_{H^\pm}^2=M_W^2+M_A^2
\quad .  
\eeq
We see that   at tree level\cite{hbound},  
 Eq. \ref{higgspot}  gives important predictions about
the relative masses of  the Higgs bosons,
\beqn
M_{H^+} &>& M_W \nonumber \\
M_H &>& M_Z \nonumber \\
M_h &<& M_A\nonumber \\
M_h &<& M_Z\mid \cos 2 \beta\mid 
 \quad .  
\label{higgmass}
\eeqn
These relations yield  the desirable
prediction that the lightest neutral
Higgs boson is lighter than the $Z$  boson and
so must be observable at LEPII.  Unfortunately
(for experimentalists at least!) it was realized 
several years ago that loop corrections to the
relations of Eq. \ref{higgmass} are large.  
In fact the corrections  to $M_h^2$ grow like $G_F M_T^4$
and receive contributions   from loops with both top
quarks 
and squarks.  In a model with unbroken supersymmetry,
these contributions would cancel. Since the supersymmetry
has been broken by splitting the masses of the
fermions and their scalar partners, 
 the neutral Higgs boson masses become
at one- loop,\cite{massloop} 
\beq
M_{h,H}^2={1\over 2}\biggl\{ M_A^2+M_Z^2+{\epsilon_h\over \sin^2
\beta}\pm\biggl[
\biggl(M_A^2-M_Z^2)\cos 2 \beta +
{\epsilon_h\over \sin^2\beta}\biggr)^2
+\biggl(M_A^2+M_Z^2\biggr)^2\sin^2 2 \beta\biggr]^{1/2}\biggr\}
\eeq
where  $\epsilon_h$ is the contribution
of the one-loop  corrections,
\beq
\epsilon_h\equiv {3 G_F\over \sqrt{2}\pi^2}M_T^4
\log \biggl( {{\tilde m}^2\over M_T^2}\biggr)
\quad .
\eeq 
We have assumed that all of  the squarks
have equal masses,  ${\tilde m}$, and have 
 neglected the smaller effects from the mixing parameters,
$A_i$ and $\mu$.  In Fig. 1, we show the lightest Higgs boson
mass
as a function of the assumed common squark mass,${\tilde m}$,
 and for two 
values of $\tan\beta$.  
For $\tan\beta > 1$, the mass eigenvalues increase monotonically
with increasing $M_A$ and give an upper bound to the
mass of the lightest Higgs boson,
\beq
M_h^2 < M_Z^2 \cos^2 2 \beta +\epsilon_h
\quad .
\eeq  
The corrections from $\epsilon_h$ are always positive and
increase the mass of the lightest neutral Higgs boson with
increasing top quark mass.  
From Fig. 1, we see that $M_h$ obtains its maximal value for
rather modest values of the pseudoscalar mass, $M_A > 300~GeV$.  
The radiative corrections to the charged Higgs mass-squared
are proportional to $M_T^2$ and so are much smaller than
the corrections to the neutral masses.

\begin{figure}[htb]
\vspace*{2.1cm}
\centerline{\epsfig{file=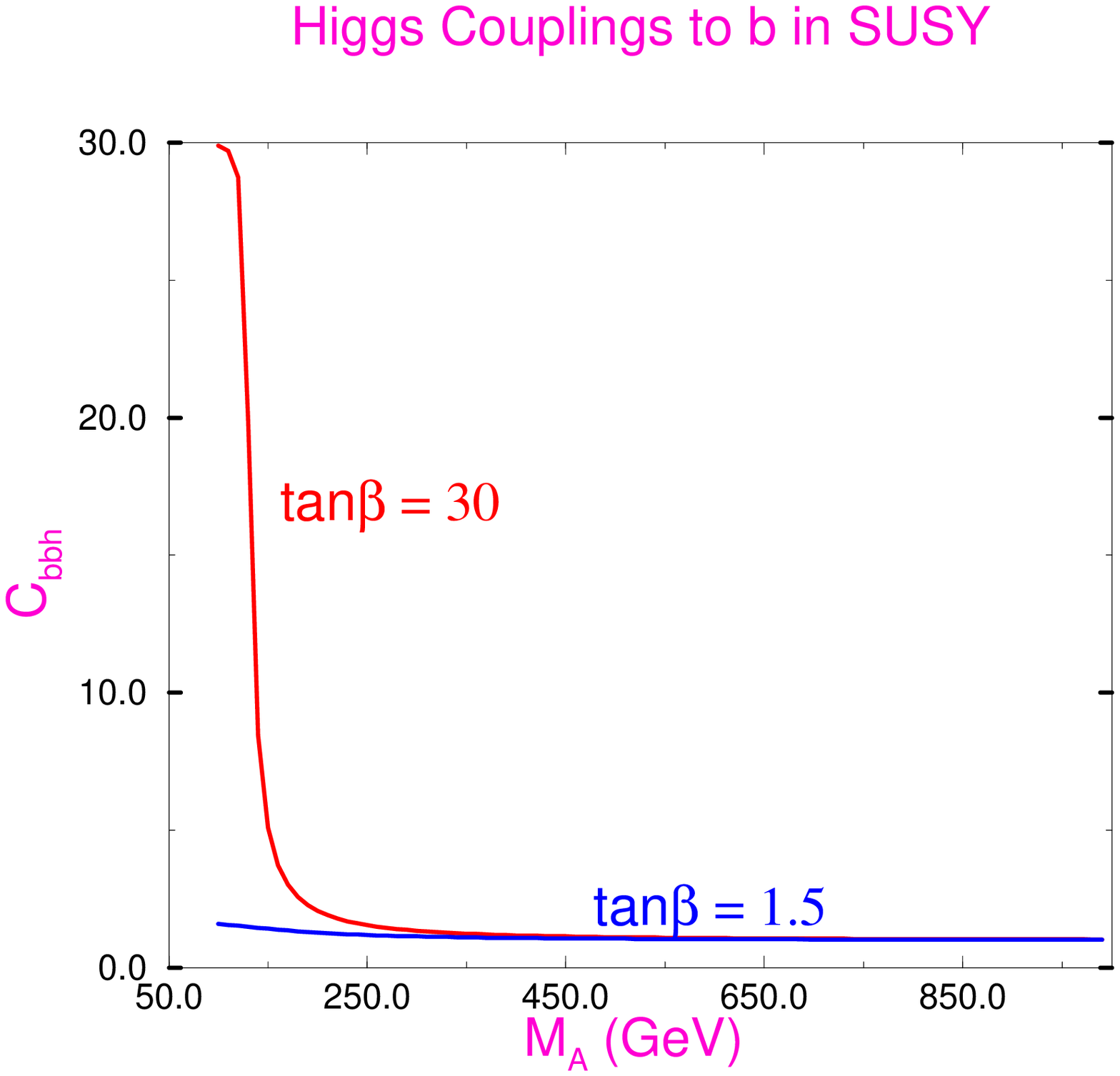,height=4.in}}
\caption{Coupling of the lightest Higgs boson
to  charge $-1/3$  quarks including radiative
corrections [28] in terms of the couplings
defined in Eq. 32. The value  
$C_{bbh}=1$ corresponds to the Standard Model coupling
of the Higgs boson to charge $-1/3$ quarks.}
\vspace*{.5in}  
\end{figure}  

There are many sophisticated analyses\cite{massloop}
which include a variety of two-loop effects, renormalization
group effects, etc., but the important point is that for
 given values  of $\tan\beta$ and the squark masses,
 there is an upper bound
on the lightest neutral Higgs boson mass.  
The maximum value of the lightest Higgs mass is shown
in Fig. 2 and we see that there is still a 
light Higgs boson even  when radiative corrections
are included.\footnote{The
leading logarithmic corrections are included in Fig. 2
and lower the result slightly from that obtained using Eq. 28.}
  For large values of $\tan\beta$
the limit is relatively insensitive to the value of
$\tan\beta$ and with  a squark mass less than about $1~TeV$,
the upper limit on the Higgs mass is about $110~GeV$.
Different approaches   
 can raise this limit
slightly to around $130~GeV$.
\begin{itemize}
\item  
The minimal SUSY model predicts a neutral Higgs
boson with a mass less than around $130~GeV$.
\end{itemize}
Such a mass scale will be accessible at LEPII or the
LHC and provides a definitive test of the MSSM.   

In a more complicated SUSY model with a richer Higgs
structure, this bound will, of course, be  changed.  However,
the requirement that the Higgs self coupling remain
perturbative up to the Planck scale gives an upper
bound on the lightest SUSY Higgs boson of around  
$150~GeV$ in all models.\cite{quiros}
This is a very strong statement.  It implies that either
there is a relatively light Higgs boson (which would
be accessible experimentally at LEPII or the LHC) or 
else there is some new physics between the weak scale
and the Planck scale which causes the Higgs
couplings to become non-perturbative.

The Higgs boson couplings to fermions
are  dictated by
the gauge invariance of the superpotential
and  at lowest order
are  completely specified in terms of the
two parameters, $M_A$ and $\tan\beta$.   
 From Eq. \ref{superpot}, we see that the charge $2/3$ quarks
get their masses entirely from $v_2$, while the
charge $-1/3$ quarks receive their masses from $v_1$.
This is a consequence of the $U(1)$ hypercharge assignments
for $H_1$ and $H_2$ given in Table 1.  In the Standard Model,
it is possible to give both the up and down quarks mass using
a single Higgs doublet.  This is because
in the Standard Model 
 the up quarks can
get their masses from the charge conjugate of the Higgs 
doublet.  Terms involving the charge conjugates of the superfields
are not allowed in SUSY models, however, and so a second Higgs
doublet with opposite $U(1)$ hypercharge
from the first Higgs doublet is necessary
in order  to give the up quarks mass.  
Requiring that the fermions have their observed masses fixes
the couplings in the superpotential
 of Eq. \ref{superpot},\cite{habergun} 
\beqn
\lambda_D &=&{g M_d\over \sqrt{2}M_W \cos\beta}\nonumber \\
\lambda_U&=&{g M_u\over \sqrt{2}M_W \sin\beta}\nonumber \\
\lambda_L &=&{g M_l\over \sqrt{2}M_W \cos\beta}
\quad ,   
\eeqn 
where $g$ is the $SU(2)_L$ gauge coupling, $g^2=4 \sqrt{2} G_F M_W^2$.  
We see that the only free parameter in the superpotential 
now is the Higgs mass parameter, $\mu$,  (along with 
the angle $\beta$
in the $\lambda_i$ couplings).

It is convenient to write  
  the couplings
for the  neutral Higgs  boson to the fermions
in terms of the Standard  Model Higgs couplings,
\beq
{\cal L}=-{g m_i\over 2 M_W} \biggl[C_{ffh}{\overline f}_i f_i h
+C_{ffH} {\overline f}_i f_i H
+C_{ffA}{\overline f}_i \gamma_5 f_i A\biggr],  
\eeq
where $C_{ffh}$ is $1$ for a Standard Model Higgs
boson.
The $C_{ffh}$ are given in Table 3 and plotted in Figs. 3 and 4
as a function of $M_A$. 
We see that for small $M_A$ and large $\tan\beta$, the couplings of
the neutral Higgs boson to fermions can be significantly different
from the Standard Model couplings; the $b$-quark coupling becomes
enhanced, while the $t$-quark coupling is suppressed.  
It is obvious from Figs. 3 and 4 that when $M_A$ becomes large
the Higgs-fermion couplings approach their standard model
values, $C_{ffh}\rightarrow 1$. In fact even for $M_A\sim 300~GeV$,
the Higgs-fermion couplings are very close to their 
Standard Model values.

\begin{table}[htb]
\begin{center}
{Table 3: Higgs Boson  Couplings to fermions}\vskip6pt
\renewcommand\arraystretch{1.2}
\begin{tabular}{|lccc|}
\hline
\multicolumn{1}{|c}{$f$}& $C_{ffh}$& $C_{ffH}$
 & $C_{ffA}  $
\\
\hline
$u$   &    ${\cos\alpha\over \sin\beta}$ &
    ${\sin\alpha\over\sin\beta}$
& $\cot\beta$ \\ 
$d$   &    $-{\sin\alpha\over\cos\beta}$ & 
     ${\cos\alpha\over\cos\beta}$ 
& $\tan\beta$  \\
\hline
\end{tabular}
\end{center}
\end{table}

\begin{figure}[htb]
\vspace*{.5in}   
\centerline{\epsfig{file=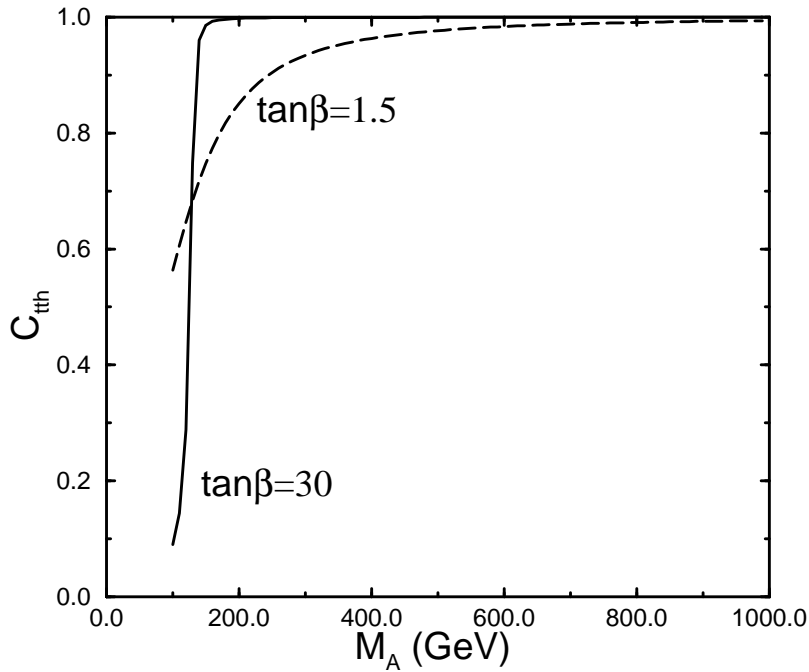,height=4.in}}
\caption{Coupling of the lightest Higgs boson
to charge $2/3$ quarks including radiative
corrections [28]  in terms of the couplings
defined in Eq. 32.
The value  
$C_{tth}=1$ yields the Standard Model coupling of the Higgs
boson  
to charge $2/3$ quarks.}
\vspace*{.5in}
\end{figure}  

\begin{figure}[htb]
\centerline{\epsfig{file=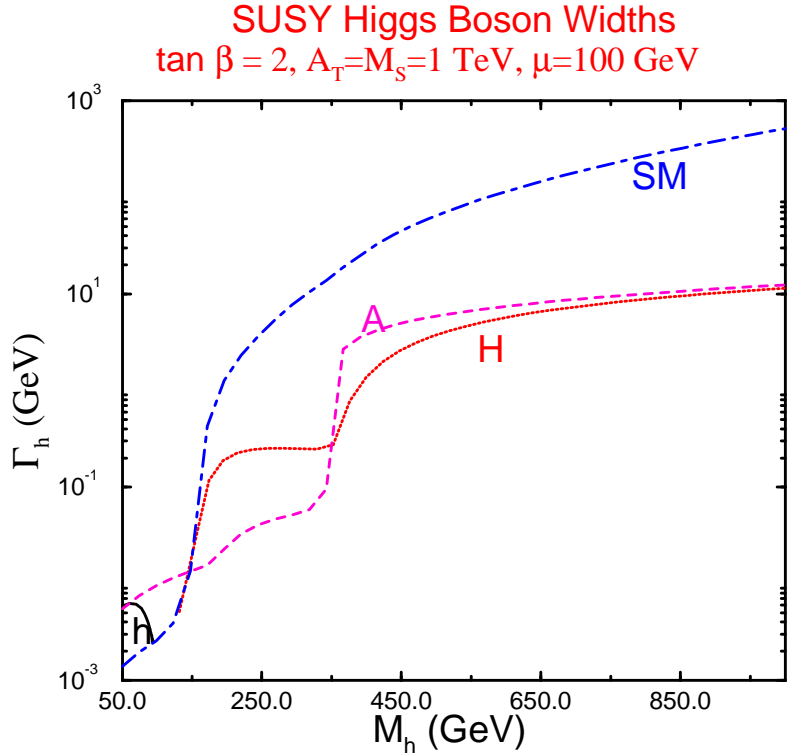,height=4.in}}
\caption{\protect Total SUSY Higgs boson decay widths including
two-loop radiative corrections as a function of the Higgs
masses. The curve for the lightest Higgs boson is cut off
at the maximum $M_h$. The program HDECAY [32] was used to obtain
this plot.}
\vspace*{.5in}  
\end{figure}  

The Higgs boson couplings to gauge bosons are fixed by the
$SU(2)_L\times U(1)$ gauge invariance.  
Some of the  phenomenologically important 
couplings  
 are:
\beqn
Z^\mu Z^\nu h:&&{igM_Z\over\cos\theta_W} \sin(\beta-\alpha)g^{\mu\nu}
\nonumber \\   
Z^\mu Z^\nu H:&&{igM_Z\over \cos\theta_W}\cos(\beta-\alpha) g^{\mu\nu}
\nonumber \\
W^\mu W^\nu h:&& igM_W \sin(\beta-\alpha)g^{\mu\nu}
\nonumber \\   
W^\mu W^\nu H:&& igM_W \cos(\beta-\alpha)g^{\mu\nu}
\nonumber \\
Z^\mu h(p)A(p^\prime):&&{g\cos(\beta-\alpha)\over 2 \cos\theta_W}
(p+p^\prime)^\mu\nonumber \\
Z^\mu H(p)A(p^\prime):&&-{g\sin(\beta-\alpha)\over 2 \cos\theta_W}
(p+p^\prime)^\mu
\quad . 
\label{vvhcoup}  
\eeqn  
We see that the couplings of the Higgs bosons to the gauge bosons
all depend on the same angular factor, $\beta-\alpha$. 
  The pseudoscalar, $A$, has no tree level coupling to pairs of
gauge bosons.   The angle $\beta$ is a
free parameter while the neutral Higgs mixing angle,
$\alpha$, which enters
into many of the couplings, can be found in terms of the physical
masses:
\beq
\tan 2 \alpha={(M_A^2+M_Z^2)\sin 2\beta
\over (M_A^2-M_Z^2)\cos 2 \beta+\epsilon_h/\sin^2\beta}
\quad . 
\eeq
With our conventions, $-{\pi\over 2}\le \alpha\le 0$.  
 It is clear that 
the couplings of the SUSY Higgs to gauge bosons are
always  suppressed
relative to those of the Standard Model.  A complete set of couplings
for the Higgs bosons (including the charged and pseudoscalar Higgs)
 at tree level 
can be found in Ref. \cite{hhg}. 
These couplings completely determine the decay modes
of the SUSY Higgs bosons and their  experimental signatures.
 The important point is that 
(at lowest order)  all of the couplings are completely determined
in terms of $M_A$ and $\tan\beta$.  When radiative corrections are
included there is a   dependence on the squark masses and 
the mixing parameters of Eq. \ref{lagsoft}.  This dependence is explored
in detail in Ref. \cite{lephiggs}.

It is an important feature  of the MSSM  that for large $M_A$,
the Higgs sector looks like that of the Standard Model.  As
$M_A\rightarrow \infty$, the masses of the charged Higgs
bosons, $H^\pm$,
and the heavier neutral Higgs, $H$, also become large leaving
only the lighter Higgs boson, $h$, in the spectrum.  In this limit,
the couplings of the lighter Higgs
boson, $h$, to fermions and gauge bosons take
on their Standard Model values. We have, 
\beqn
\sin(\beta-\alpha)&& 
\rightarrow 1~ {\hbox {for}}~M_A\rightarrow \infty
\nonumber \\ 
\cos(\beta-\alpha)&&\rightarrow  0 
\quad . 
\eeqn
From Eq. \ref{vvhcoup}, we see that the heavier Higgs
boson, $H$, decouples from the gauge bosons in the heavy $M_A$ limit,
while
the lighter Higgs boson, $h$, has Standard Model couplings.
 Figs. 3 and 4 demonstrate  that
the Standard Model limit is also
 rapidly approached in the fermion-Higgs couplings for $M_A > 300~GeV$.
  In the
limit of large $M_A$, it will thus 
 be exceedingly difficult to differentiate a SUSY
Higgs sector from the  Standard Model Higgs boson.
\begin{itemize}
\item
The SUSY Higgs sector with large $M_A$ looks like the
Standard Model Higgs sector.
\end{itemize}                                 

The total width of the Higgs boson depends sensitively on $\tan\beta$
and is illustrated in Fig. 5 for $\tan\beta=2$.\cite{squrad}
  We see that
the lightest Higgs boson has a width $\Gamma_h\sim 10-100~MeV$,
while the heavier Higgs boson has a width $\Gamma_H\sim .1-1~GeV$,
which is considerably narrower than the width of the Standard Model
Higgs boson with the same mass.  (The curve for the lighter Higgs
boson is cut off at the kinematic upper limit.)  
The pseudoscalar, $A$, is also narrower than a Standard
Model Higgs boson with the same mass.

\subsection{The Squark and Slepton Sector}

We turn now to a discussion of the scalar partners of the quarks
and leptons.  The left-handed $SU(2)_L$ quark doublet has
scalar partners,
\beq
{\tilde Q}= 
\left( \begin{array}{c}
 {\tilde u}_L\\
{\tilde d}_L \end{array}\right)
\quad .
\eeq
The right-handed quarks also have scalar partners, ${\tilde u}_R$
and ${\tilde d}_R$.  The L and R subscripts denote which  helicity quark
the scalars are partners of--
{\bf they are for identification 
purposes only.
These are ordinary complex scalars}.  Before SUSY is broken
the fermions and scalars have the same masses and this mass
degeneracy is split by the soft mass terms of Eq. \ref{lagsoft}.
The tri-linear $A$ terms allow the scalar partners of the
left- and right-handed fermions to mix to form the mass
eigenstates.  
  In the
top squark sector, the mixing between the scalar
partners of the left- and right handed top (the stops), ${\tilde t}_L$
and ${\tilde t}_R$,  is given by
\beq
M_{{\tilde t}}^2=\left(
\begin{array}{ll}
{\tilde M}_Q^2+M_T^2+M_Z^2({1\over 2}-{2\over 3}
\sin^2\theta_W)\cos 2 \beta & M_T(A_T+\mu \cot\beta)\\
M_T(A_T+\mu\cot\beta)& {\tilde M}_U^2+M_T^2
+{2\over 3}M_Z^2\sin^2\theta_W \cos 2 \beta
\end{array}\right)
\quad .
\label{squarkm}
\eeq 
For the scalars associated with the lighter
quarks, the mixing effects will be negligible,
since the mixing is proportional to the quark mass, 
(except if $\tan\beta >> 1$, when  ${\tilde b_L}-
{\tilde b_R}$ mixing may be large).
   
From Eq. \ref{squarkm}, we see that there are two important cases
to consider.  If the soft breaking occurs at a large scale,
much greater than $M_Z$, $M_T$,  and $A_T$,
then  all the soft masses
will be   approximately
equal, and we will have $12$ degenerate squarks with
mass ${\tilde m}\sim {\tilde M}_Q
\sim {\tilde M}_U\sim
{\tilde M}_D$.  On the other hand, if the soft masses
and the tri-linear mixing term, $A_T$,  
are on the order of the electroweak scale, then mixing
effects become important.  
  
\begin{figure}[htb]
\vspace*{1.in}  
\centerline{\epsfig{file=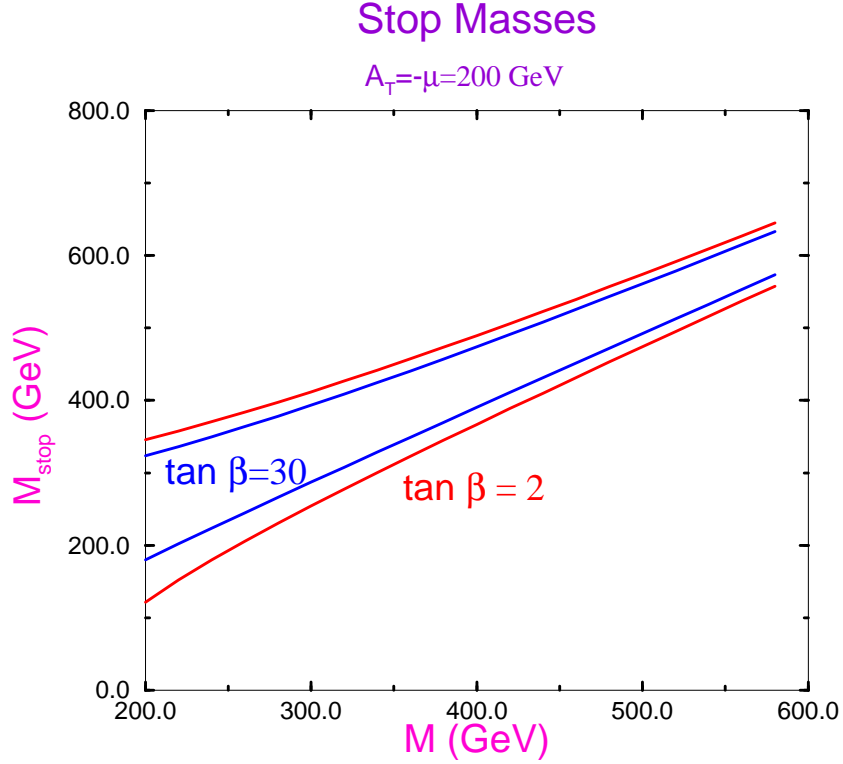,height=4.in}}
\caption{Stop squark masses for large mixing parameters,
$A_T=\mu=200~GeV$, and for $\tan\beta=2$
and $\tan\beta=30$.  $M\equiv {\tilde M}_Q={\tilde M}_u$ are  the
squark mass parameters of Eq. 37.} 
\vspace*{.5in} 
\end{figure}  
If mixing effects are large, then  one of the stop squarks will become
the lightest squark, since the mixing effects are proportional
to the relevant quark masses and hence will be largest in this
sector.
The case where the lightest squark is the stop is particularly interesting
phenomenologically, and we discuss it in the section on squark mass
limits.\cite{stop}  
In Fig. 6, we show the stop squark masses
for ${\tilde M}_Q={\tilde M_t}=
{\tilde M_b}\equiv {\tilde m}$ and
for several values of $\tan\beta$.
  Of course the mixing effects 
cannot be too large, or the stop squark mass-squared will
be driven negative, leading to a breaking of the 
color  $SU(3)$ gauge symmetry.  
Typically, the requirement that the correct vacuum be chosen
leads to a restriction on the mixing parameter on the order
of $\mid A_T\mid <  {\tilde m}$.\cite{bagtasi}

The couplings of the squarks to gauge bosons are
completely  fixed by 
gauge invariance, with no free parameters.   A few
examples of the couplings are:
\beqn
\gamma^\mu ~{\tilde q}_{L,R}(p)~{\tilde q}_{L,R}^*(p^\prime):   
&&-i e Q_q (p+p^\prime)^\mu \nonumber \\
W^{\mu -} ~ {\tilde u}_L(p) ~ {\tilde d}_L^*(p^\prime): 
&&-{i g \over \sqrt{2}}(p +p^\prime)^\mu \nonumber \\
Z^\mu ~ {\tilde q}_{L,R}(p) ~ {\tilde q}_{L,R}^*(p^\prime) : 
&&
-{i g \over \cos\theta_W} \biggl[T_3 - Q_q \sin^2
\theta_W \biggr] (p + p^\prime)^\mu
\quad , 
\eeqn  
where $T_3$ and $Q_q$ are the quantum numbers of 
the corresponding quark.
The strength of the interactions are clearly given by the
relevant gauge coupling           constants.
  A complete set of Feynman rules can
be found in Ref. \cite{hkrep}  

The mixing in the slepton sector is analogous to that in
the squark sector and we will not pursue it further.    
From Table 1, we see that the scalar partner of the $\nu_L$,
${\tilde \nu}_L$, has the same gauge quantum numbers
 as the $H_2^0$ 
Higgs boson.  It is possible to give ${\tilde \nu}_L$ a 
vacuum expectation value and use it to break the electroweak
symmetry.  Such a vacuum expectation value would break lepton
number (and $R$ parity) 
thereby giving the neutrinos a mass and so its magnitude
is severely restricted. \cite{rparity}

\subsection{The Chargino Sector}
  
There are two charge $1$, spin- ${1\over 2}$ Majorana
fermions; ${\tilde \omega}^\pm$, the fermion partners
of the $W^\pm$ bosons, and ${\tilde h}^\pm$, the charged
fermion partners of the Higgs boson, termed the Higgsinos.
The physical mass states, ${\tilde \chi}_{1,2}^\pm$, are
linear combinations formed by diagonalizing the mass
matrix and are usually called charginos.  In the 
${\tilde \omega}^\pm - {\tilde h}^\pm$ basis the
chargino mass matrix is,
\beq 
M_{\tilde \chi^\pm}=
\left(\begin{array}{cc}
M_2 & \sqrt{2}M_W\sin \beta \\
\sqrt{2}M_W\cos\beta &  - \mu
\end{array}
\right)
\quad .  
\eeq
The physics is extremely sensitive to $M_2/\mu$.  
The mass eigenstates are then,
\beq
M^2_{{\tilde \chi}^\pm_{1,2}}={1\over 2}\biggl\{
M_2^2+2 M_W^2+\mu^2\mp
\biggl[ (M_2^2-\mu^2)^2+4M_W^4\cos^2 2 \beta
+ 4M_W^2(M_2^2+\mu^2-2M_2\mu\sin^2\beta
\biggr]^{1/2}\biggr\}  .
\eeq 
By convention $M_{{\tilde \chi}_1^\pm}$ is the
lighter chargino.

\subsection{The Neutralino Sector} 
  
In the neutral fermion sector, the
neutral fermion partners of
the $B$ and $W^3$ gauge bosons, ${\tilde b}$ and
${\tilde \omega}^3$, can mix with the neutral
fermion partners of the Higgs bosons, ${\tilde h}_1^0,
{\tilde h}_2^0$.  Hence the physical states,
${\tilde \chi}_i^0$,  are
found by diagonalizing the $4\times 4$ mass
matrix, 
\beq
M_{\tilde \chi_i^0} = 
\left(\begin{array}{cccc}
M_1 & 0& -M_Z \cos\beta \sin\theta_W & M_Z \sin\beta\sin\theta_W\\
0 & M_2 & M_Z \cos\beta\cos\theta_W & -M_Z \sin\beta\cos\theta_W\\
-M_Z \cos\beta\sin\theta_W & M_Z \cos\beta\sin\theta_W & 0 & \mu \\
M_Z \sin\beta\sin\theta_W & -M_Z \sin\beta\cos\theta_W & \mu & 0 
\end{array}
\right)
\eeq
where $\theta_W$ is the electroweak mixing angle
and we work in the ${\tilde b}, {\tilde \omega}^3,
{\tilde h}_1^0, {\tilde h}_2^0$ basis.  The physical
masses can be defined to be positive and by convention,
$M_{{\tilde \chi}_1^0}
< M_{{\tilde \chi}_2^0}
< M_{{\tilde \chi}_3^0}
< M_{{\tilde \chi}_4^0}$.
In general, the mass eigenstates do not correspond to
a photino, (a fermion partner of the photon), or a 
zino, (a fermion partner of the $Z$), but are complicated
mixtures of the states. The photino is only a mass eigenstate
if $M_1=M_2$.  Physics involving the neutralinos therefore
depends on $M_1$, $M_2$, $\mu$, and
$\tan\beta$.  
The lightest neutralino, ${\tilde \chi}_1^0$, is usually assumed
to be the LSP.  
 
\section{WHY DO WE NEED SUSY?}
 Having introduced the MSSM as an effective
theory at the electroweak scale  and briefly
discussed the various new particles and interactions  of the model, I
turn now to a discussion of the reasons for constructing 
a SUSY theory in the first place.
  We have already discussed the cancellation
of the quadratic divergences, which is automatic in
a supersymmetric model.   
There are, however, many other reasons why theorists are excited about
supersymmetry.
Theorists will often state that the mathematics of a supersymmetric
model is ${\it beautiful}$.  However, in my mind, the beauty
of supersymmetry is largely
obscured by the ugliness of the SUSY breaking sector which we
have introduced, and  it is therefore  essential to have a solid
motivation for studying SUSY theories.  
\subsection{Coupling constants run!}

In a gauge theory, coupling constants scale with energy
according to the relevant $\beta$-function.  Hence, having
measured a coupling constant at one energy scale, its value at any
other energy can be predicted.  At  one loop, 
\beq
{1\over \alpha_i(Q)}={1\over \alpha_i(M)}+
{b_i\over 2 \pi} \log\biggl({M\over Q}\biggr)
\quad .
\label{smbeta}
\eeq
In the Standard (non-supersymmetric) Model, the coefficients
$b_i$ are given by,
\beqn
b_1&=& {4\over 3}N_g+{N_H\over 10}\nonumber \\
b_2&=& -{22\over 3}+{4\over 3}N_g+{N_H\over 6}
\nonumber \\
b_3&=&-11+{4\over 3}N_g \quad ,
\eeqn
where $N_g=3$ is the number of generations and $N_H=1$ is
the number of Higgs doublets. 
The evolution of the coupling constants is seen to be sensitive to
the particle content of the theory.  
  We can take $M=M_Z$ in Eq. \ref{smbeta}, input the
measured values of the coupling constants at the $Z$-pole and
evolve the couplings to high energy.  The result is shown in Fig. 7.  
There is obviously no meeting of the coupling constants at high
energy.  
  
\begin{figure}[htb]
\vspace*{1.in}  
\centerline{\epsfig{file=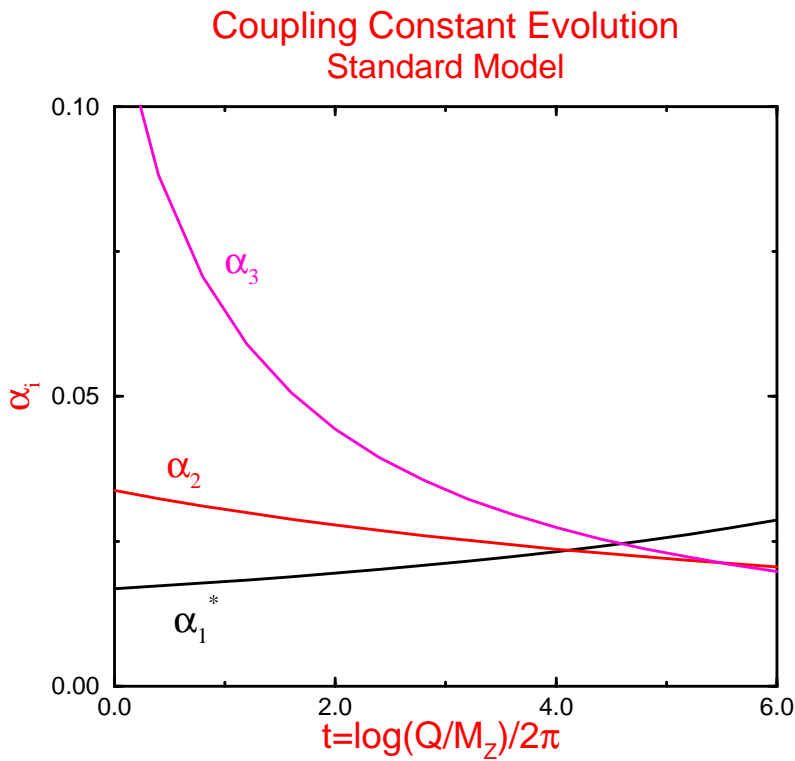,height=4.in}}
\caption{Evolution of the gauge coupling
constants in the Standard Model from the experimentally
measured values at the $Z$-pole.  
$\alpha_1^*\equiv 5/3 \alpha_1$, since this is the relevant
coupling in Grand Unified Theories. }
\vspace*{.5in}  
\end{figure}

\begin{figure}[htb]
\vspace*{1.in}    
\centerline{\epsfig{file=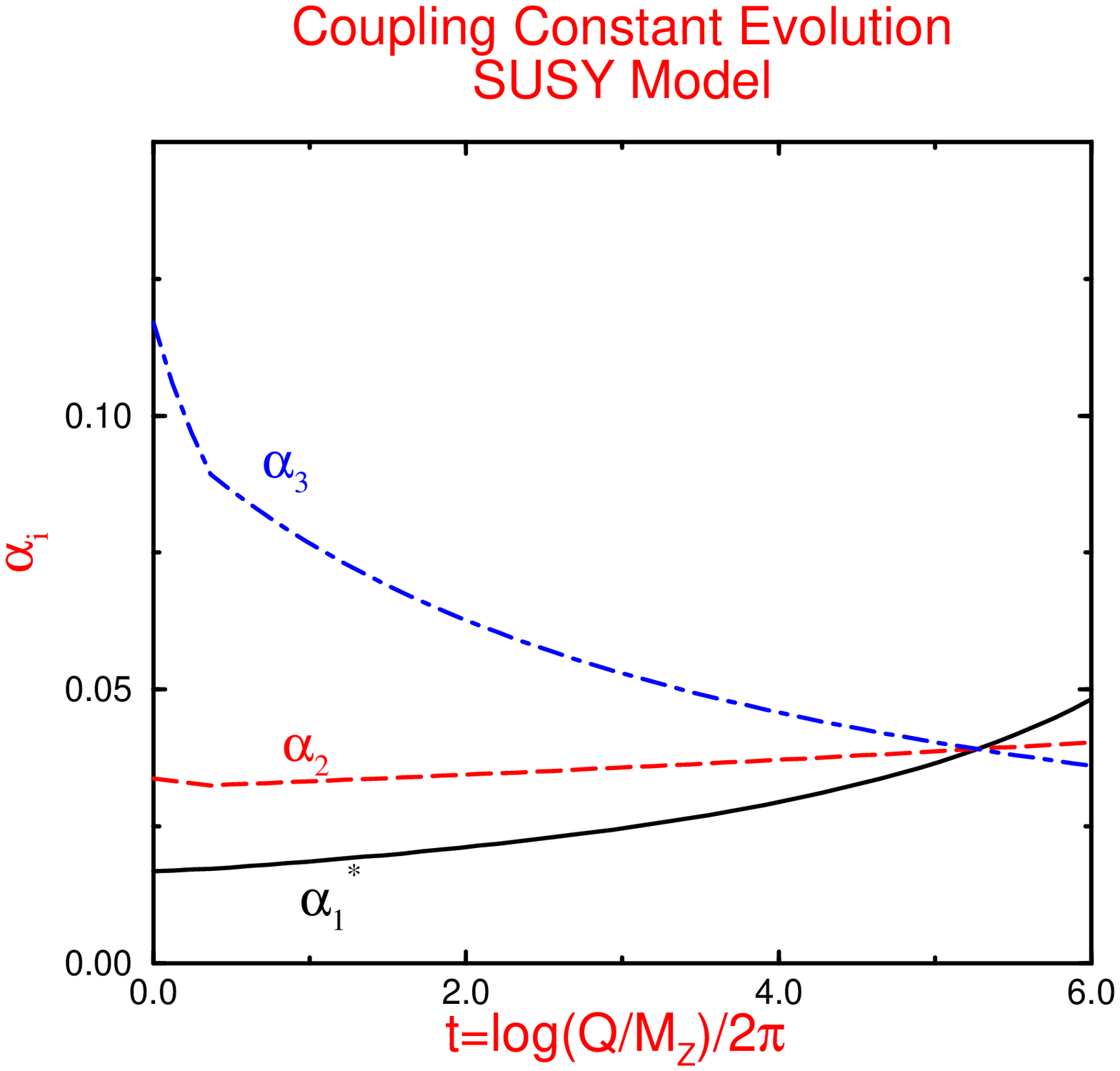,height=4.in}}
\caption{Evolution of the coupling constants
in a low energy SUSY model
from the experimentally measured values at the $Z$-pole.
The SUSY thresholds are taken to be at
$1~TeV$. 
$\alpha_1^*\equiv 5/3 \alpha_1$, since this is the
relevant coupling in Grand Unified Theories.  }  
\vspace*{.5in}  
\end{figure}

If the theory is supersymmetric, then the spectrum is different
and the new particles contribute to the evolution of the coupling constants.
In this case we have,\cite{betafuns} 
\beqn
b_1&=& 2 N_g+{3\over 10}N_H\nonumber \\
b_2&=& -6+2 N_g+{N_h\over 2}\nonumber \\
b_3&=& -9+ 2 N_g
\quad .
\eeqn
Because a SUSY model of necessity contains two
Higgs doublets, we have $N_H=2$.  If we assume that the mass of
all the SUSY particles is around $1~TeV$, then the coupling
constants scale as shown in Fig. 8.  We see that the coupling
constants meet at a scale around $10^{16}$ GeV.\cite{early,unif, ccunif} 
This meeting of the coupling constants is a necessary feature of 
a Grand Unified Theory (GUT).  

\begin{itemize}
\item
SUSY theories can be naturally incorporated
into Grand Unified Theories.
\end{itemize}
There are many variations on this theme including  two
loop beta functions, effects from passing
through SUSY particle thresholds, etc.,
but they all allow us to take the picture of SUSY as
resulting from  a GUT theory 
seriously.\cite{ccunif,jb}  
  
\subsection{SUSY GUTS}

The observation that the measured coupling constants tend to meet at
a point when evolved to high energy 
assuming the $\beta$-function of a low energy SUSY model
has led to widespread acceptance of
a standard SUSY GUT model.  We assume that the 
$SU(3)\times SU(2)_L\times U(1)$ gauge coupling
constants are unified at a high scale $M_X\sim 10^{16}~GeV$:\footnote{
This normalization of the $U(1)_Y$ coupling constant is canonical
in Grand Unified Theories.} 
\beq
\sqrt{{5\over 3}}g_1(M_X)=g_2(M_X)=g_3(M_X)\equiv g_X
\quad .
\eeq
The gaugino masses, $ M_i$, are also assumed to unify,
\beq  M_i(M_X)\equiv m_{1/2}
\quad .
\eeq
At lowest order, the gaugino masses  then scale in the same
way as the corresponding coupling constants,
\beq
M_i(M_W)=m_{1/2}{g_i^2(M_W)\over g_X^2}
\eeq
yielding  
\beqn
 M_2&=& {\alpha\over \sin^2\theta_W}{1\over
\alpha_s}  M_3\nonumber \\
 M_1&=& {5\over 3} \tan^2\theta_W  M_2
\quad .
\eeqn
The gluino mass is always the heaviest of the
gaugino masses.  
This relationship between the gaugino masses is a fairly robust
prediction of SUSY GUTS and persists in  models where
the supersymmetry is broken dynamically.\cite{snow,mess}  
 
Typical SUSY GUTS also assume that there is a common
scalar mass at $M_X$,
\beqn
&m_1^2(M_X)=&m_2^2(M_X)\equiv m_0^2\nonumber \\
{\tilde M}_Q^2(M_X)= {\tilde M}_d^2(M_X)=& 
{\tilde M}_u(M_X)=&{\tilde M}_L^2(M_X)=
{\tilde M}_e^2(M_X)\equiv m_0^2
\quad .
\eeqn  
The neutral Higgs boson
masses at $M_X$ are then $M_{h,H}^2=m_0^2+\mu^2$.
As a final simplifying assumption, a common $A$ parameter is
assumed,
\beq
A_T(M_X)=A_b(M_X)=....\equiv A_0
\quad .
\eeq
With these assumptions, the SUSY sector is completely described
by 5 input parameters at the GUT scale,\cite{fp}
\begin{enumerate}
\item
A common scalar mass, $m_0$.
\item
A common gaugino mass, $m_{1/2}$.
\item
A common trilinear coupling, $A_0$.
\item
A Higgs mass parameter, $\mu$.
\item
A Higgs mixing parameter, $B$.
\end{enumerate} 
This set of assumptions is often called the ``superstring
inspired SUSY GUT" or SUGRA (although the connection with
superstrings  and/or supergravity
is mostly wishful thinking) or the
``constrained MSSM" (CMSSM).  Although this framework is somewhat
${\it ad~hoc}$, it does provide guidance to reduce the immense
parameter space of a SUSY model.
In actual practice, these relationships are satisfied
only in the simplest models.  

  The strategy is now to 
input the $5$ parameters given above   at $M_X$  and to  use the
renormalization group equations to evolve the parameters to $M_W$. 
In fact, the requirement that the $Z$ boson obtain its measured
value  when the parameters are evaluated   at 
low energy can be used to restrict $\mid \mu B\mid$, leaving
the $sign(\mu)$ as a free parameter.  We can also trade
the parameter $B$ for $\tan\beta$.  In this way the parameters
of the model 
become
\beq
m_0, m_{1/2}, A_0, \tan\beta, {\rm sign}(\mu)
\quad .
\eeq
  This
form of a SUSY theory is extremely predictive, as
the entire low energy spectrum is predicted in terms of a 
few input parameters.
Within this scenario, contours for the
various SUSY particle masses can be found as a function
of $m_0$ and $m_{1/2}$ for given values of $\tan\beta$, 
$A_0$ and ${\rm sign}(\mu)$.\cite{jb,fp}
 
It is instructive to study the scalar masses within this
scenario.  
  The evolution of the sleptons between $M_X$ and $M_W$
 is small and
we have the approximate result
for the slepton masses,\cite{xerxes,jb}  
\beq
{\tilde M}_L(M_W)^2\sim {\tilde M}_e(M_W)^2\sim m_0^2,
\eeq
while the squark masses are roughly
\beq
{\tilde M}_q^2 (M_W)\sim m_0^2+4 m_{1/2}^2
\quad .
\eeq  
Since the squarks 
have strong interactions, (which drives the masses upwards),
their masses at the weak scale tend to be larger than the sleptons.  
Once all the particle masses have been computed in this
scheme, then their production cross sections and
decay rates at any given accelerator can be computed
unambiguously.

  Changing the input parameters at 
$M_X$ (for example, assuming non-universal scalar masses)
of course changes the phenomenology at the weak scale.  A
preliminary investigation of the sensitivity of the low energy
predictions to these assumptions has been made in Ref. \cite{snow}.
For now, we will consider the Grand Unified Model described
above as a starting point for phenomenological investigations
into SUSY and hope that the general search strategies 
developed for this model will be applicable to other models.   

\subsection{Electroweak Symmetry Breaking}

The simple SUSY model described above has the appealing feature
that it explains the mechanism of electroweak symmetry breaking.
Below, we sketch the argument.

In the Standard Model (non-supersymmetric) with a single Higgs field, $\phi$,
the scalar potential is given by:
\beq
V(\phi)=\mu^2 \phi^2+\lambda \phi^4 
\quad .
\eeq
By convention, $\lambda>0$.  If $\mu^2>0$, then $V(\phi)>0$
for all $\phi$ not equal to $0$
 and there is no electroweak symmetry breaking.
If, however, $\mu^2<0$, then the minimum of the potential is not
at $\phi=0$ and the potential has the familiar Mexican hat shape.
When the Lagrangian is expressed in terms of the physical field,
$\phi^\prime\equiv \phi-v$, which
has zero vacuum expectation value, then
the electroweak symmetry is broken and the $W$ and $Z$ gauge bosons
acquire non-zero masses. 
We saw  in the previous sections
that this same mechanism gives the $W$ and $Z$ gauge
bosons their masses in the MSSM.
 This simple picture leaves one looming question: 
\beq
 {\rm Why~ is~ 
\mu^2< 0? } 
\nonumber
\eeq
  It is this question which the  SUSY GUT  models can
answer.

In the minimal SUGRA model which we have described above,  the neutral Higgs
bosons both 
have masses, $M_{h,H}^2=m_0^2+\mu^2$,
at $M_X$  while the squarks and sleptons have
mass $m_0$ at $M_X$.   Clearly, at $M_X$, the
electroweak symmetry is not broken since the Higgs
 bosons have positive mass-squared.
 The masses scale with energy according to the 
renormalization group equations.\cite{yukren}
  If we neglect gauge couplings and 
consider only the scaling of the third generation scalars
we have,\cite{ewsbtop}
\beq
{d\over d\log(Q)}\left(\begin{array}{c}
M_h^2\\
{\tilde M}_{t_R}^2\\
{\tilde M}_{Q_L^3}^2\end{array}\right)=-{8 \alpha_s\over 
3 \pi}  M_3^2
\left(\begin{array}{c}
0\\
1\\
1\end{array}\right)
+{\lambda_T^2\over 8 \pi^2}
\biggl({\tilde M}_{Q_L^3}^2+{\tilde M}_{t_R}^2
+M_h^2+A_T^2\biggr)
\left(\begin{array}{c}
3\\
2\\
1\end{array}\right) \quad ,
\label{scaling}
\eeq
where ${\tilde Q}_{L}^3$ is the $SU(2)_L$ doublet containing
${\tilde t}_L$ and ${\tilde b}_L$, $h$ is the lightest Higgs
boson, $\lambda_T$ is the top quark Yukawa coupling
constant given in Eq. 31,  
 and $Q$ is the effective scale at which
the masses are measured.  The signs are such that the Yukawa
interactions (proportional to $M_T$) decrease the masses, while the
gaugino interactions increase the masses.  Because of the $3-2-1$ structure of
the last term in Eq. \ref{scaling},
 the Higgs mass decreases faster than the
squark masses and it is possible to drive $M_h^2<0$ at low energy,
while keeping ${\tilde M}_{Q_L^3}^2$ and
${\tilde M}_{t_R}^2$ positive. 
A generic set of scalar masses in a typical SUSY GUT model is shown
in Fig. 9.  We can clearly see that the lightest Higgs boson
mass becomes negative around the electroweak scale.\cite{masssamp}    

 For large $\lambda_T$, we have
the approximate solution,
\beq
M_h^2(Q)=M_h^2(M_X)-{3\over 8\pi^2}\lambda_T^2
({\tilde M}^2_{Q_L^3}+{\tilde M}_{t_R}^2
+M_h^2+A_T^2)\log\biggl({M_X\over Q}\biggr)
\quad .
\eeq
Hence the larger $M_T$ is, the faster $M_h^2$ goes negative.
This of course generates electroweak symmetry breaking.
If $M_T$ were light, $M_h^2$ would remain positive.\cite{ewsbtop} 
This observation was made ten years ago when we thought
the top quark was light, ($\sim 40~GeV$).  At that
time it was ignored as not being phenomenologically relevant.
In fact, this mechanism only works for $M_T\sim 175~GeV$!  
\begin{itemize}
\item
SUSY GUTS can explain electroweak symmetry breaking.
The lightest Higgs boson mass is negative,
$m_h^2<0$, because $M_T$ is large.
\end{itemize}
  
\begin{figure}[htb]
\centerline{\epsfig{file=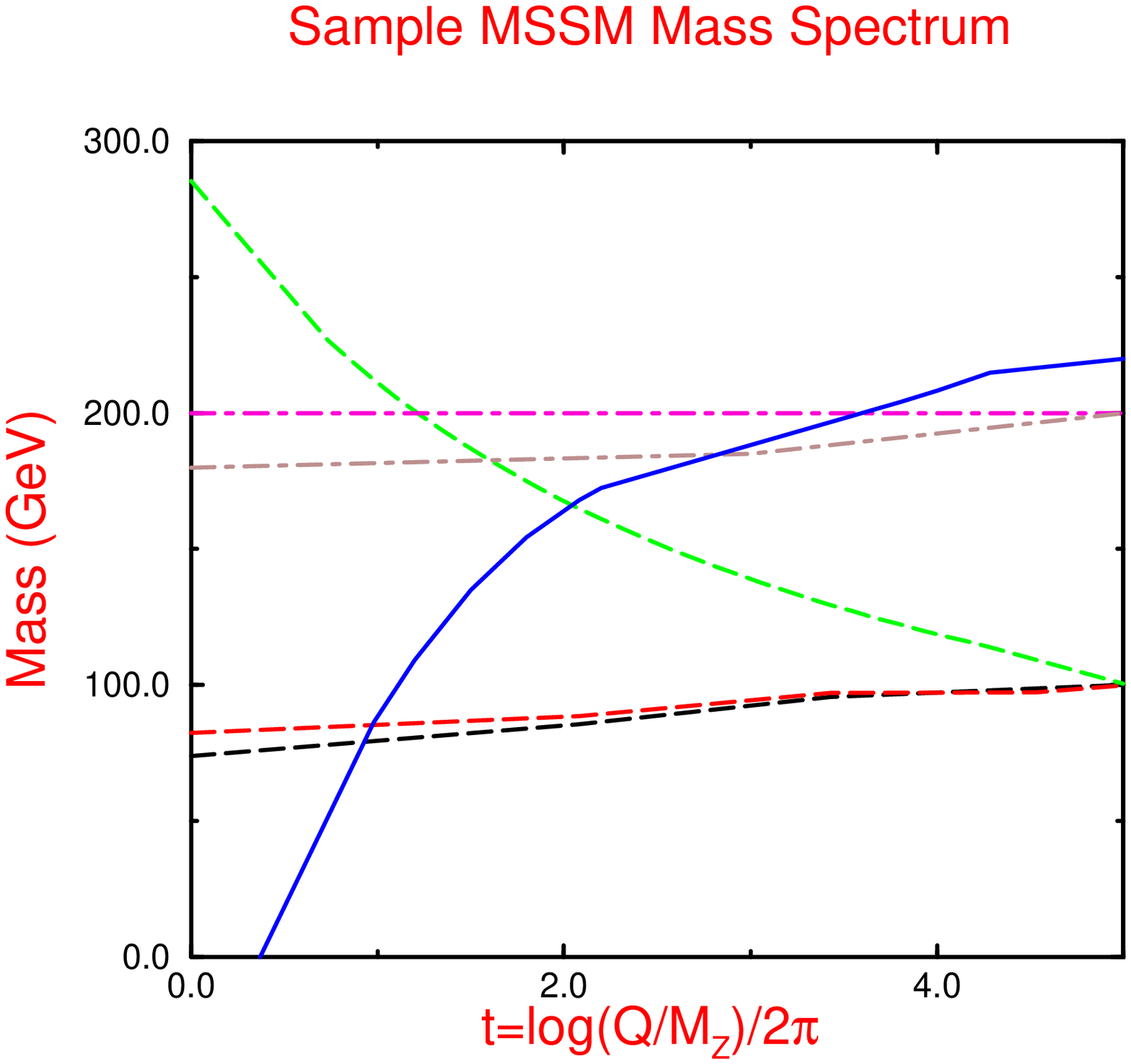,height=4.in}}
\caption{Sample masses of SUSY particles in a SUSY GUT.
At the GUT scale $M_X$, we have taken $m_0=200~GeV,
m_{1/2}=100~GeV, \mu=100~GeV$ and $A_i=0$.  The solid
line is the lightest neutral Higgs boson mass.  The dashed
lines are the gaugino masses (the largest is the gluino) and the
dot-dashed lines are typical squark masses. }  
\vspace*{.5in}   
\end{figure}  

The $3-2-1$ structure of Eq. \ref{scaling} drives
$M_h^2$ negative faster than the squark masses.  This
is important because driving the squark mass negative would
have the undesired effect  of breaking the
color $SU(3)$ symmetry. 
The requirement that the electroweak symmetry breaking occur
through the renormalization group scaling of the Higgs boson
mass, (as given in Eq. \ref{scaling}) also restricts the
allowed values of $\tan\beta$ to $\tan \beta > 1$. 
(Remember that $\lambda_T$ depends on $\beta$ through Eq. 31.)

\subsection{Fixed Point Interactions}

In the previous subsection we saw that a large top
quark mass could generate electroweak symmetry breaking
in a SUSY GUT  model.  Here we show
that the simplest SUSY GUT actually ${\it predicts}$
a large top quark mass.

The top quark mass is determined in terms of its Yukawa
coupling and scales with energy, $Q$,\cite{btau} 
\beq
\lambda_T(Q)={M_T(Q)\over M_W}
{g\over \sqrt{2}\sin\beta}
\quad .
\eeq
Including both the gauge couplings and the Yukawa couplings
to the $t$- and $b$- quarks, the scaling is:
\beq
{d \lambda_T\over d \log(Q)}=
{\lambda_T\over 16 \pi^2}
\biggl\{
-{13\over 9} g_1^2-3 g_2^2-{16\over 3} g_3^2
+6 \lambda_T^2+\lambda_B^2\biggr\}
\quad .
\eeq
To a good approximation, we can consider only the contributions
from the strong coupling constant, $g_3$, and the top
quark Yukawa coupling,
$\lambda_T$.  If we begin our scaling at 
$M_X$ and evolve $\lambda_T$ to lower energy, we will come to a
point where the evolution of the Yukawa coupling stops,
\beq
{d \lambda_T\over d \log(Q)}=0
\quad .
\eeq
At this point we have roughly, 
\beq
-{16\over 3} g_3^2+6 \lambda_T^2=0
\eeq
which gives,
\beq
\lambda_T\sim{4\over 3}\sqrt{2 \pi\alpha_s}\sim 1
,
\eeq
or
\beq
M_T\sim (200~GeV) \sin\beta \quad . 
\eeq
This point where the top quark mass stops evolving is called
a ${\it fixed~point}$.
What this means is that no matter what the initial condition
for $\lambda_T$ is at $M_X$, it will always evolve to
give the same value at low energy.  
  For $\tan\beta\sim2$, the fixed point
value for the top quark mass is close to the experimental value.  
More sophisticated analyses do not change this picture substantially.  
\begin{itemize}
\item
SUSY GUTS can naturally accommodate a large top quark mass
for $\tan\beta\sim 1-3.$
 
\end{itemize}

\subsection{$b-\tau$ Unification}

The unification of the $b$- and $\tau$- Yukawa coupling
constants,
$\lambda_B$ and $\lambda_\tau$, 
at the GUT scale is a concept much beloved by
theorists since
\beq
\lambda_B(M_X)=\lambda_\tau(M_X)
\label{btau} 
\eeq
occurs naturally in many GUT models.  Requiring that
the $b$ quark have its experimental value  at low energy leads to
a prediction for the top quark mass in terms of $\tan\beta$.
There are two solutions which yield $M_T=175~GeV$,\cite{btau}
\beqn
\tan\beta&\sim& 1\nonumber \\
{\rm or}\quad \tan\beta&\sim& {M_T\over M_b}
\quad .  
\eeqn 
The first solution roughly corresponds to the fixed point solution
of the previous subsection. 
The second solution with $\tan\beta\sim 35$  has interesting
phenomenological consequences, since for large $\tan\beta$
the coupling of the lightest Higgs boson to $b$ quarks is enhanced
relative to the Standard Model. (See Fig. 3). 
 The  values in the $\tan\beta -
M_T$ plane
allowed by $b-\tau$ unification
 depend sensitively on the exact value of the strong coupling
constant, $\alpha_s$, used in the evolution and so there is a
significant uncertainly in the prediction.    
\begin{itemize}
\item
SUSY GUTs allow for 
the unification of the $b-\tau$
Yukawa coupling constants at the GUT
scale  along with the experimentally
observed value for the top quark mass.
\end{itemize}
Similar relationships to Eq. \ref{btau} involving the first two generations
do not work.  
\subsection{Comments}
We see that SUSY plus grand unification has many desirable
features and can explain a lot:
\begin{enumerate}
\item 
There are no troubling  quadratic divergences requiring
disagreeable cancellations.
\item
$M_T$ is large because $\lambda_T$ evolves from the GUT
scale to its fixed point.
\item
Electroweak symmetry is broken, $m_h^2<0$, because $M_T$ is
large. 
\item
$b-\tau$ unification can be incorporated, leading to the 
experimentally observed value for the top quark mass.
\end{enumerate} 
Afficianados of SUSY can add many more items to this list.\cite{topten} 
For instance, the LSP is a leading candidate for
cold, dark matter.\cite{dark} 
  The
conclusion is inescapable:
                
 ${\centerline{{\bf SUSY~IS ~HERE~TO~
STAY !}}}$

\section{SEARCHING FOR  SUSY}

We begin this section with a description of the effects of
SUSY particles on precision measurements and rare decays.
We then turn to experimental limits on the various particles
and search strategies at current and future machines.
A more detailed expose can be found in the lectures of
Tata\cite{xerxes}
along with up to the minute limits in Refs. \cite{mer,sch}.  
 
\subsection{Indirect Hints for SUSY}

One might hope that the precision measurements at the
$Z$-pole could be used to garner information on the SUSY
particle spectrum.  Since the precision electroweak measurements
are overwhelmingly in good agreement with the predictions of
the Standard Model, it would appear that stringent
limits could be placed on the existence of SUSY particles at
the weak scale.  There are two reasons why this is not the 
case.

  The first is that SUSY is a ${\it decoupling~theory}$.
With the exception of the Higgs particles,
 the effects of SUSY particles at the weak scale are
suppressed by powers of $M_W^2/M_{SUSY}^2$, where $M_{SUSY}$ is
the relevant SUSY mass scale,  and so
for $M_{SUSY}$ larger than a few hundred $GeV$,
the SUSY particles  give negligible contributions to 
electroweak processes.
The second reason why there are not stringent limits
from precision results at LEP has to do with the Higgs 
sector.  The Higgs bosons are the only particles
in the spectrum  which  do not decouple
from the low energy physics when they are very massive.
The
fits to electroweak data tend to prefer a Higgs boson in
the $100~GeV$ mass  range.\cite{dpf}  Since the MSSM requires a light
Higgs boson with a mass in this region
anyways,   the electroweak data is completely consistent
with a SUSY model with a light Higgs boson and all other SUSY
particles significantly heavier.

Attempts have been made to perform global fits to the electroweak
data and to fix the SUSY spectrum this way.\cite{deb,ewsusy}
It is possible to obtain a fit where the $\chi^2$/degree of
freedom is roughly the same as in the Standard Model fit. 
Although the fits do not yield stringent limits
on the SUSY particle masses, they do    
exhibit several interesting features.
  They tend to prefer either small $\tan\beta$, $\tan\beta \sim 2$, or
 relatively large values, $\tan\beta\sim 30$.  
In addition, the fitted values for the strong coupling constant
at $M_Z$,
$\alpha_s(M_Z)$,  are slightly smaller in  SUSY  models than
in the Standard Model.  (For $\tan\beta=1.6$, Ref. \cite{deb}
finds $\alpha_s(M_Z)=.116\pm.005$ and for $\tan\beta=34$, they
find $\alpha_s(M_Z)=.119\pm.005$.)   
It is clear that all precision electroweak measurements can
be accommodated within a SUSY model, but the data show
no preference for  these models.   

There are also numerous indirect limits coming from the
effects of SUSY particles on rare decays.  
Since the SUSY particles circulate in loops, they can affect
rare $B$ and $K$ decays (among others).
One
of the most restrictive  limits is from
the CLEO measurement of the
inclusive decay  $B\rightarrow X_s\gamma$,\cite{cleo} 
\beq
BR(B\rightarrow X_s\gamma)=(2.32\pm .67)\times 10^{-4}
, \qquad {\rm CLEO}  
\eeq  which
is sensitive to
loops containing the new particles of a SUSY model.
The contribution from $tH^\pm$ loops always  adds constructively
to the Standard Model result and hence non-
supersymmetric two- Higgs doublet 
models are severely restricted by the measurement of $b\rightarrow s\gamma$.
  
The situation is different in a SUSY model, however, since there 
are additional contributions from squark-chargino loops, squark-
neutralino loops, and squark-gluino loops.  The contributions
from the squark-neutralino and squark-gluino loops are small and
are typically neglected.  The 
dominant contribution from the squark-chargino
loops  is proportional to $A_T\mu$
 and thus 
can have either sign relative to  the Standard Model and
charged Higgs loop contributions.
  There will therefore   be regions of SUSY parameter
space which are excluded depending upon whether there is
constructive or destructive interference between the Standard
Model/ charged Higgs contributions and the squark-chargino
contribution.\cite{bsg}   
 The limit which  can be obtained is obviously   very sensitive to the
sign$(A_T\mu)$ and can be easily understood  for large
$\tan\beta$ where the squark-chargino contribution
is  completely dominant.    Neglecting QCD corrections (which
are significant) we have,\cite{bsg2}
\beq  
{BR(b\rightarrow s\gamma)\over BR(b\rightarrow c e {\overline \nu})}
\sim{\mid V_{ts}V_{tb}\mid^2\over \mid V_{cb}\mid^2}
{6 \alpha\over \pi}
\biggl\{
C+{M_T^2 A_T\mu\over {\tilde m}_T^4}\tan\beta\biggr\}^2
\quad ,
\eeq
where $C$ (positive) is the contribution from the Standard
Model and charged Higgs loops and ${\tilde m}_T$ is the stop mass.
For $A_T\mu$ positive, this leads to a larger branching
ratio, $BR(b\rightarrow s\gamma)$, than in the Standard Model.
Since the Standard Model prediction is already somewhat
above the measured value,
we require $A_T\mu<0$ to avoid conflict
with the experimental measurement 
if ${\tilde m}_T$ is at the electroweak scale 
and $\tan\beta$ is large.
\footnote{Reader beware:  There are conflicting definitions
of the sign of $\mu$ in the literature.  The only way to
be sure is to go back to the superpotential of Eq. \ref{superpot} to
see how $\mu$ is defined. 
} 
 Detailed plots of the allowed regions for various 
assumptions about $\tan\beta$, $\mu$, and $A_T$ are given in Ref.
\cite{bb}.  
Depending on $\tan\beta$ 
and the sign of $A_T\mu$,
this process probes stop masses in the $100-300~GeV$
region. 
For large $\tan\beta$, $B\rightarrow X_s \gamma$ may probe
mass scales as large as a $TeV$.\cite{jhjw} 
 
Another class of important indirect limits on SUSY models comes
from flavor changing neutral current (FCNC)  processes
such as $K^0-  {\overline K}^0$ mixing.
In general, the matrix which diagonalizes the squark mass
matrix is different from that which diagonalizes the quark
mass matrix and so there are off-diagonal interactions
which can mediate FCNC's.  
  The contributions
from squarks to FCNC processes vanish if the squarks have 
degenerate masses and so the limits are typically of the form:
\beq
{\Delta{\tilde m}^2\over {\tilde m}^2}< {\cal O}(10^{-3})
\qquad ,  
\eeq
where $\Delta {\tilde m}^2$ is the mass-squared splitting
between the different squarks and ${\tilde m}$ is the
average squark mass.  
A detailed discussion of FCNCs in SUSY models and references
to the literature is given in Ref. \cite{fcnc}.   As a practical
matter, the assumption is often made that there are $10$ degenerate
squarks, corresponding to the scalar partners of the $u,d,c,s,$ and
$b$ quarks, while the stop squarks  are allowed to have different 
masses from the others.
  This avoids phenomenological problems with FCNCs.    
\section{Experimental Limits and Search Strategies}

We turn now  to  a discussion of some of the
existing experimental limits  on
the various SUSY particles and also to the search 
strategies applicable at present and future accelerators.
This section is intended only to give the flavor of how
SUSY searches proceed and not as a comprehensive guide.
We begin with the Higgs sector.
\subsection{Observing SUSY Higgs Bosons}

The goal in the Higgs sector is to observe the
$5$ physical Higgs particles, $h,H,A,H^\pm$, and to measure as
many couplings as possible to verify that the couplings are those
of a SUSY model.
  The lightest neutral Higgs boson in 
the minimal SUSY model is unique in the SUSY spectrum because there
is  an upper bound to its mass,
\beq
M_h<130~GeV.
\eeq
All other SUSY particles in the model
can be made arbitrarily heavy just by adjusting the soft
SUSY breaking parameters in the model and so be
just out of reach of today's or
tomorrow's accelerators (although if they
are heavier than around $1~TeV$, much of the  motivation
for low energy SUSY disappears). The lightest SUSY Higgs
boson, however, cannot be much outside the range of LEPII
and can almost certainly be observed at the LHC .  Hence an
extraordinary theoretical effort has gone into the 
study of the reach of various accelerators in the SUSY
Higgs parameter
space since in this sector it will be possible to
experimentally exclude the MSSM if a light Higgs boson is
not observed.

 If we find
a light neutral Higgs boson, then we want to map out the
parameter space to see if we can distinguish it
from a Standard Model Higgs boson.
The only way to do this is to measure a variety of production
and decay modes and attempt to extract the various
couplings of the Higgs bosons to fermions and gauge
bosons.  
Since as $M_A\rightarrow \infty$, the $h$ couplings approach
those of the standard model, there will clearly be a region
where the SUSY Higgs boson  and the Standard Model Higgs boson are
indistinguishable.   
This is obvious from Figs. 3 and 4.

The search strategies for the SUSY Higgs boson depend
sensitively on the Higgs  boson branching ratios, which
in turn depend on $\tan\beta$.  In Figs. 10 and 11,
we show the branching ratios for the lightest SUSY 
Higgs boson, $h$, into some interesting
decay modes assuming that there are no SUSY
particles light enough for the $h$ to decay into.
(These figures include radiative corrections to
the branching ratios, which can be important.\cite{squrad})  
For a Higgs boson below the $WW$ threshold, the
decay into $b {\overline b}$ is  completely dominant.  Unfortunately,
there are large QCD backgrounds to this decay mode and so it
is often necessary to look at rare decay modes. 
The branching ratios to $b {\overline b}$, $\tau^+\tau^-$,
and $\mu^+\mu^-$ are relatively insensitive to $\tan\beta$,
but the $W W^*$, $ZZ^*$, and $\gamma\gamma$ rates have
 strong dependences on $\tan\beta$ as we can see from
 Figs. 10 and 11.  

\begin{figure}[htb]
\vspace*{1.in}  
\centerline{\epsfig{file=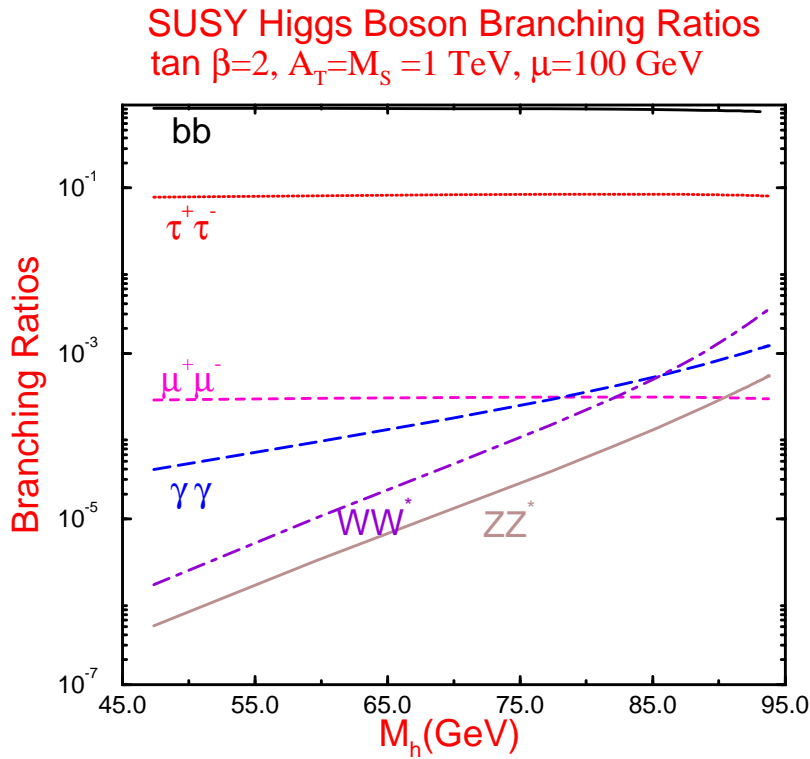,height=4.in}}
\caption{Branching ratios of the lightest Higgs boson
assuming  decays into other SUSY particles are
kinematically forbidden.  $WW^*$ and $ZZ^*$ denote
decays with one off-shell gauge boson and $M_S$ is a typical squark
mass.[32]}
\vspace{.5in} 
\end{figure}  

\begin{figure}[htb]
\vspace*{1.in}  
\centerline{\epsfig{file=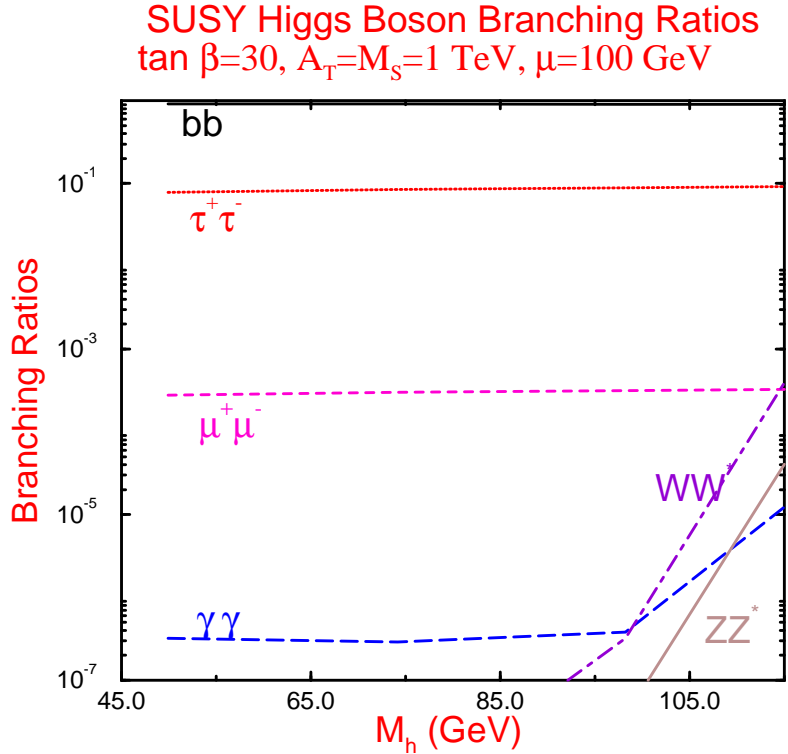,height=4.in}}
\caption{Branching ratios of the lightest Higgs boson assuming
decays into other SUSY particles are  kinematically forbidden.[32]}
\vspace*{.5in}  
\end{figure} 

Direct limits on SUSY Higgs production have
been obtained at LEP by searching for the complementary
processes,\cite{lephiggs}
\beqn
e^+e^- &\rightarrow & Z h\nonumber \\
e^+e^-&\rightarrow & Ah
\quad .  
\eeqn 
From the couplings of Eq. 33, we see that the process $e^+e^-\rightarrow
Zh$ is suppressed by  $\sin^2(\beta-\alpha)$ relative
to the Standard Model Higgs boson production process, while
$e^+e^-\rightarrow Ah$  is proportional to  $\cos^2(\beta -\alpha)$.
  The moral is
that it is impossible to suppress both processes simultaneously
if both the $h$ and the $A$ are kinematically accessible!  
The  experimental searches  look for final states with $b$'s and $\tau$'s  
since these have the largest branching ratios. 
Because the Higgs sector can be described by the two parameters,
$M_h$ and $\tan\beta$, searches exclude a region in this plane.
 (Remember that $M_A$ can be expressed in terms of
$M_h$ and $\tan\beta$ at lowest order.  When radiative corrections 
are included, there will be a dependence on the mixing
parameters, $A_i$ and $\mu$, and on
the squark masses).  
The LEP searches for Higgs bosons, 
$e^+e^-\rightarrow Zh$ and $e^+e^-\rightarrow AH$,  exclude
the region, \cite{lephiggs} 
\beq
M_h> 44~GeV, ~{\rm for~any~}\tan\beta
\quad .
\eeq
For a given value of $\tan\beta$, there may be a stronger bound.
It is important to note that the LEP searches do not leave any 
window for a very  light (on the order of a few
$GeV$) Higgs boson.
  The limit on a SUSY Higgs boson is weaker than the 
corresponding limit on the  Standard Model Higgs boson,
$M_h^{SM}>65~GeV$,
  due to the suppression in the couplings of the Higgs boson
to vector bosons.     

\begin{figure}[htp]
\vspace*{1.in} 
\centerline{\epsfig{file=eezh.epsi}}
\vspace*{1.5in}  
\caption{Cross sections for $e^+e^-\rightarrow
Zh$ and $e^+e^-\rightarrow A h$ at LEP.
From Ref. [31].} 
\end{figure}

At LEPII, the cross section for either $Zh$ (small
$\tan\beta$) or $Ah$ (large $\tan\beta$) is roughly
$.5~pb$.  With a luminosity of $150/pb/yr$, this leads to
$75$ events/yr before the inclusion of branching ratios.
 Fig. 12 shows the cross sections for two different
values of $\tan\beta$ and the complementarity of the 
two processes can be clearly observed.  (The dependence
on the top quark mass arises from the inclusion of
radiative corrections.)

\begin{figure}[htb]
\vspace*{1.in}  
\centerline{\epsfig{file=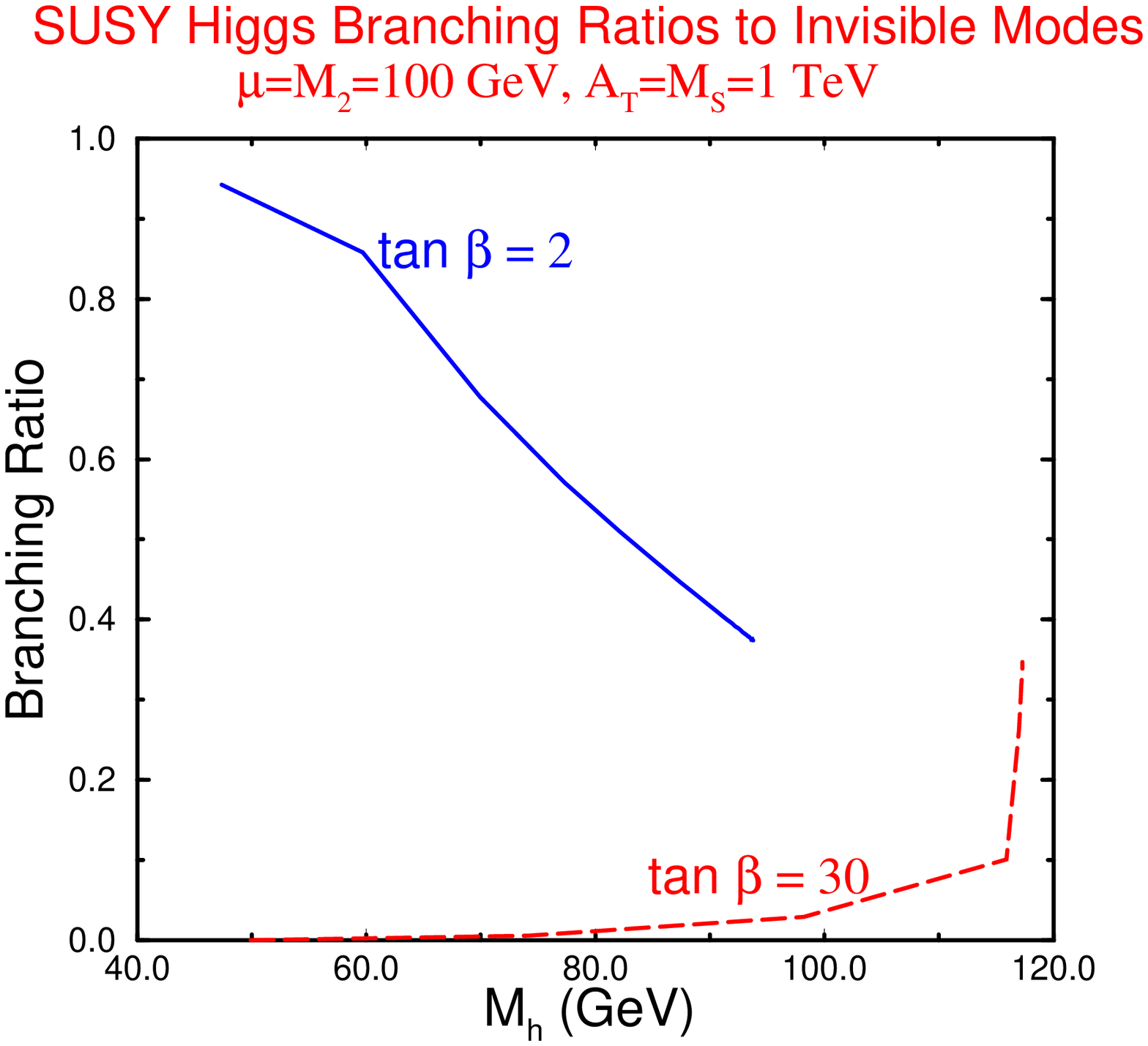,height=4.in}}
\caption{Branching ratio of the lightest Higgs boson
to ${\tilde \chi}_1^0{\tilde \chi}_1^0$.
The curve with $\tan\beta=30$ has $M_{\chi_1^0}=33~GeV$,
while that with $\tan\beta=2$ has $M_{\chi_1^0}=7~GeV$.[32]}
\vspace*{.5in}  
\end{figure} 
The limits on the Higgs boson mass
 could be substantially altered if there is a significant
branching rate into invisible decay modes, such as the
neutralinos, 
\beq h,A\rightarrow
{\tilde \chi}_1^0{\tilde \chi}_1^0
\quad .
\eeq
These branching ratios  could be as high as $80\%$, but
are  extremely
model dependent since they depend sensitively on 
the parameters of the neutralino mixing matrix.
In Fig. 13, we show the branching ratio of the lightest
Higgs boson to ${\tilde \chi}_1^0{\tilde \chi}_1^0$ for
several choices of parameters.  For
$\tan\beta=2$, with the set of parameters which we have chosen, the
branching ratio is always greater than $40\%$.
  If the invisible decay modes are significant,
  a different search strategy  for the Higgs boson must
be utilized and LEPII can put a limit on the 
product of the Higgs boson mixing angles, $\beta-\alpha$, and
the branching ratio to invisible modes:
\beqn
R_1^2&\equiv&\sin^2(\beta-\alpha)BR(h\rightarrow {\hbox{visible}})
\nonumber \\
R_2^2&\equiv& \sin^2(\beta-\alpha
)BR(h\rightarrow {\hbox{invisible}})
\quad .
\eeqn
For $M_h=40~GeV$, the $95\%$ confidence level excluded region
from LEP is,\cite{lephiggs}
\beqn
R_1^2&<&.3\nonumber \\
R_2^2&<&.1
\quad .
\eeqn
These limits can  be reinterpreted in terms
of the parameters of the MSSM ($A_i$, $\mu$, $M_2$, $M_1$,
${\tilde m}$, etc.) and will 
 be greatly improved at LEPII.  For
$M_h=80~GeV$ and an integrated luminosity of $150/pb$ at $\sqrt{s}=192~GeV$,
the $95\%$ confidence level limit will be:
\beqn
R_1^2&<&.1
\nonumber \\
R_2^2&<&.3
\quad .
\eeqn
These limits will significantly restrict the allowed
SUSY parameter space. 

A $\mu^+\mu^-$ collider could in principle obtain stringent bounds
on a SUSY Higgs boson 
through its $s$-channel couplings to the Higgs.\cite{bargermm}
Since these couplings are proportional to the lepton mass,
the $s$-channel Higgs couplings will be much larger
at a $\mu^+\mu^-$ collider than at an $e^+e^-$ collider.   
For large $\tan\beta$, the lighter Higgs boson could be found in
the process $e^+e^-\rightarrow Zh$
 at LEPII or at an NLC.\cite{lephiggs,nlchiggs}
 However, for   
large $\tan\beta$, the coupling of the heavier Higgs boson to gauge
boson pairs is highly suppressed, (see Eq. 33),
 so the $H$ can't be
found through $e^+e^-\rightarrow ZH$.  Instead the $H$ can be found
through $\mu^+\mu^-\rightarrow H\rightarrow b {\overline b}$, which
is enhanced by the factor $\tan^2\beta$ relative to
$\mu^+\mu^-\rightarrow h_{SM}\rightarrow b {\overline b}$.

A muon collider could also  be very useful for
obtaining precision measurements of the lighter Higgs boson mass.
The idea is that the $h$ has been discovered through
either the process $e^+e^-\rightarrow Z h$ or $\mu^+
\mu^-\rightarrow Z h$ and so we have a rough idea of the
Higgs boson mass.
A muon collider could  be tuned to sit right on
the resonance, $\mu^+\mu^-\rightarrow h$.
  By doing an energy scan around the region
of the resonance, a precise value of the mass could be 
obtained due in large part to the narrowness of the muon beam
as compared to the beam in an electron collider.  (The
narrowness of the beam is
due to the suppression of synchrotron radiation in a muon collider.)

 At the LHC,  for most Higgs masses the dominant
production mechanism is gluon fusion, $gg\rightarrow h,H$ or $A$.
These processes proceed through triangle diagrams with
internal $b$ and $t$
quarks and also through squark loops.  In the limit in which the
top quark is much heavier than the Higgs boson, the top
quark contribution is a constant, while the $b$ quark contribution
is suppressed by $(M_b/v)^2\log(M_h/M_b)$ and
so only the top quark contribution is numerically
important.  For large $\tan\beta$,
however, the dominance of the top quark loop is overtaken by the 
large ${\overline b}
bh$ coupling and the bottom quark contribution becomes
important, (as seen in Fig. 3).
The production rate is therefore extremely sensitive to $\tan
\beta$.  
  Both QCD corrections and squark loops can also be
numerically important.\cite{spira} In fact, the QCD
corrections increase the rate by a factor between $1.5$ and $2$. 
  The rate for $pp\rightarrow h$
at the LHC is
shown in Fig. 14 as a function of $\tan\beta$ for $M_h=80~GeV$.
  We see that there are a relatively large number
of events.  For example, for $M_h\sim 80~GeV$, the LHC cross section is
roughly $100~pb$.  With a luminosity of $10^{33}/cm^2/sec$, this
yields $10^6$ events/year.

\begin{figure}[htb]
\vspace*{1.in}  
\centerline{\epsfig{file=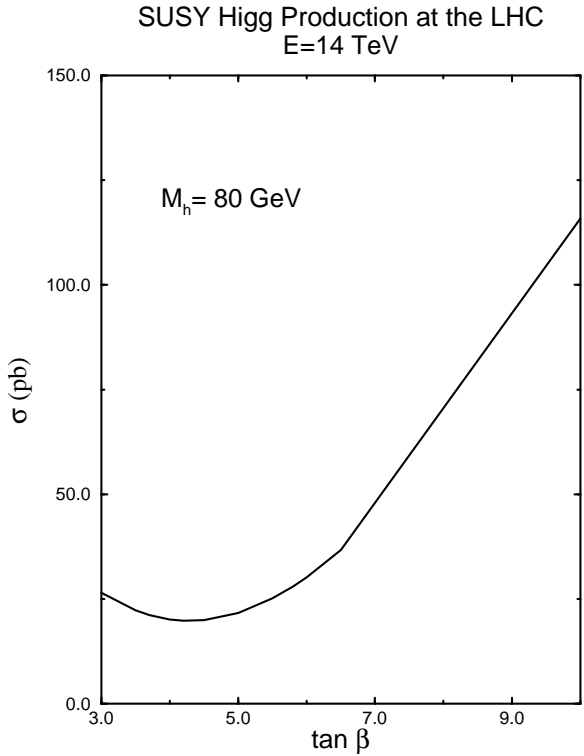,height=4.in}}
\caption{Cross section for production of the lightest
SUSY Higgs boson at the LHC as a function of $\tan\beta$.}    
\vspace*{.5in}  
\end{figure}  

Unfortunately, there are large backgrounds to the dominant decay
modes, 
( $b \overline{b}, \mu^+\mu^-$, and $\tau^+\tau^-$),
for a Higgs boson in the $100~GeV$ region.\cite{lhchiggs}
The decay $h \rightarrow 
Z Z^*$ will be useful, but its
branching ratio decreases rapidly with decreasing
Higgs mass.  In order to cover the region around $M_h\sim 80-100~GeV$,
 it will be 
necessary to look for the Higgs decay to $\gamma\gamma$,
\beq
gg\rightarrow h,H\rightarrow \gamma\gamma
\quad .
\eeq
(From Figs. 10 and 11, we see that the $BR(h\rightarrow
\gamma\gamma)$ is typically $< 10^{-3} - 10^{-5}$.)
This process will be extremely difficult to observe at
the LHC due to the small rate and the desire to observe
the $h\rightarrow \gamma\gamma$  decay
has been one of the driving forces behind the design
of both LHC detectors.\cite{lhcprop}
For large
$M_A$, the rate is roughly independent of $\tan\beta$
for $\tan\beta>3$ 
and can be used to exclude $M_A>150~GeV$ with
the full design luminosity of $3\times 10^{5}/pb$.
(With a smaller luminosity of $3\times 10^4/{\rm pb}$,
the $h\rightarrow \gamma\gamma$ process is sensitive to
roughly $M_A>270~GeV$.  See Fig. 15 for the exact region.)

In order to exclude the region with smaller $\tan\beta$, the 
process $pp\rightarrow Wh\rightarrow l \nu {\overline b} b$
can be used.\cite{wm}
  This process can exclude a region with $M_A > 
100~GeV$ and $\tan\beta < 4$ (see Fig. 15)
 and demonstrates the crucial need for $b$-tagging
at the LHC in order to cover all  regions
of SUSY parameter space.
In Fig. 15, we see the excluded region formed by combining
the LHC and LEP limits.\cite{dfr} 
A variety of Higgs production and decay channels can be
utilized in order to probe the entire $\tan\beta-M_A$ plane.
 The most striking feature of Fig. 15 is the
region
 around $M_A\sim100~GeV$ for $\tan\beta > 5$ where 
the lightest Higgs boson cannot be observed.
In the region with
$M_A\sim 100-200~GeV$, both the $h {\overline t} t$ coupling
and the $h\rightarrow \gamma\gamma$ branching ratios are suppressed
relative to the Standard Model rates.  Furthermore, the dominant
decays, $h\rightarrow b {\overline b}$ and $h\rightarrow 
\tau^+\tau^-$, have large backgrounds from $Z$ decays.
It will be necessary to look for the decays of the heavier
neutral Higgs boson, $H$, or the pseudoscalar, $A$,
 to $\tau^+\tau^-$
pairs in order to probe this region,
\beq
H,A\rightarrow \tau^+\tau^-\rightarrow l \nu {\overline q} q
\quad .
\eeq
Detector studies by the ATLAS and CMS collaborations suggest
that these decay modes may be accessible.

\begin{figure}[p]
\centerline{\epsfig{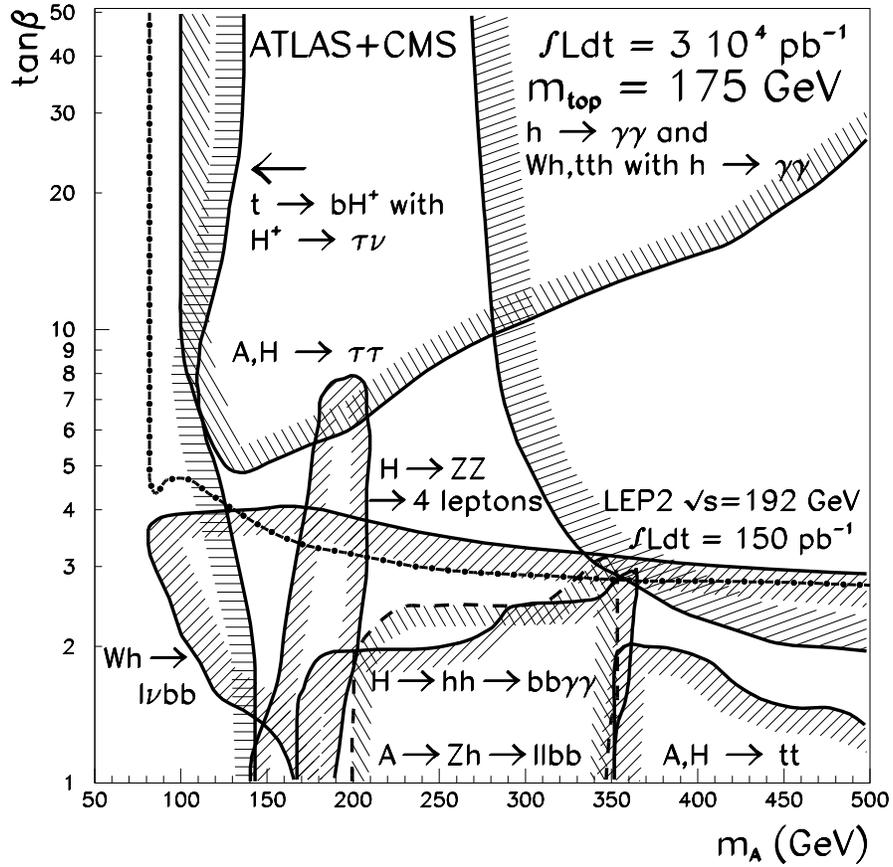}} 
\vspace*{4.5in}  
\caption{LHC (with low luminosity)
  and LEPII discovery limits for SUSY
Higgs bosons. Figure from Ref. [60].}
\vspace*{.5in}  
\end{figure}

\section{Finding the Zoo of SUSY Particles}

In addition to the multiple Higgs particles associated
 with SUSY models, there is a whole zoo of other new
particles.  There are the squarks and gluinos which
are produced through the strong interactions and the sleptons,
charginos, and neutralinos which are produced weakly.

We begin by discussing some generic signals for supersymmetry.
All SUSY particles 
in a theory with $R$ parity conservation
eventually decay to the LSP, which is
typically taken to be the lightest neutralino, ${\tilde \chi}_1^0$, 
although in some models it could be the gravitino.\cite{wein}
  The LSP's interactions with matter are extremely
 weak and so it escapes detection leading to missing energy.  
 
\begin{itemize}
\item
A basic SUSY signature is missing energy, $E_T^{miss}$,
 from the undetected LSP.
\end{itemize}
A SUSY model typically produces a cascade of decays, until the final state
consists of only the LSP plus jets and leptons.  Hence typical final 
states are:
\begin{itemize}
\item
$l^\pm+{\rm jets}+E_T^{miss}$
\item
$l^\pm l^\pm + {\rm jets}+E_T^{miss}
$
\item $l^\pm l^\mp+{\rm jets}+E_T^{miss}$ 
\quad .
\end{itemize} 
Because of the presence of the LSP in the final state, it is not possible to
completely reconstruct the masses of the SUSY particles, although
a significant amount of information about the masses
can be obtained from the event structure.
\begin{itemize}
\item
A combination of characteristic signatures
 may determine the SUSY model.
\end{itemize}

Because the gluinos are Majorana particles, they have some special
characteristics which may be useful for their experimental detection.
They have the property:
\beq
\Gamma({\tilde g}\rightarrow l^+ X)=
\Gamma({\tilde g}\rightarrow l^- X)\quad .
\eeq
Hence gluino pair production can lead to final states with same sign 
$l^\pm l^\pm$ pairs.\cite{fp, ssll}
  The standard model background for this
type of signature is rather small.
\begin{itemize}
\item 
Same sign di-lepton pairs are a useful signature for gluino pair
production.
\end{itemize} 

Another generic signature for SUSY particles is tri-lepton 
production.\cite{tri}  If 
we consider the process of chargino-neutralino production,then it is 
possible to have the process:
\beq
{\tilde \chi}_1^\pm {\tilde \chi}_2^0\rightarrow
l \nu {\tilde\chi}_1^0+
{\overline l}^\prime l^\prime {\tilde \chi}_1^0
\quad .
\eeq
Again this is a signature with a small standard model background.

In the following sections we will examine several of these signatures 
in detail.
In order to predict the SUSY particle production rates,
it is necessary to have an event generator which includes both
the production and decays of the SUSY particles.  A number of
generators exist for both $e^+ e^-$ and hadronic colliders.
The physics assumptions of
  two of the most commonly used event generators for SUSY
(ISASUSY and SPYTHIA)  are
reviewed in Ref. \cite{eventgen}.

\subsection{Chargino and Neutralino Production}

As an example of SUSY particle searches, we consider the
search for chargino pair production at an electron-positron
collider,
\beq
e^+ e^- \rightarrow {\tilde \chi}_1^+ {\tilde \chi}_1^-
\quad , 
\eeq 
(where ${\tilde \chi}_1^\pm$ are the lightest charginos.)
The chargino mass matrix has a contribution 
from both  the fermionic partner
of the $W^\pm$,
 ${\tilde\omega}^\pm$, and from the fermionic partner of the
charged
Higgs, ${\tilde h}^\pm$, and so depends on the two unknown parameters
in the mass matrix, $\mu$ and $M_2$.  (See Eq. 40).
If $\mid \mu \mid << M_2$, we say the chargino is ``Higgsino-like",
while if $\mid \mu \mid >> M_2$, it is termed ``gaugino-like".
Results are usually presented in terms of the mass of the lightest chargino, 
$M_{\tilde \chi^+_1}$, and $\mu$.  

There are two types of
Feynman
 diagrams contributing to chargino pair production: 
the first is an $s$-channel exchange of a $\gamma$ or a $Z$, and the
second is the $t$-channel
exchange of  the scalar partner of the neutrino,
 ${\tilde \nu}_L$.  There is a destructive interference
between the two types of diagrams.  The largest interference occurs
for light ${\tilde \nu}_L$ and ${\tilde \chi}^\pm_1$ ``Gaugino-like".
For  light ${\tilde \nu}_L$, ${\tilde m}_{ \nu_L}< 60~GeV$,  
 the destructive interference  can make the cross
section  significantly  smaller, leading to a weaker limit.    
For a heavy ${\tilde\nu}_L$,
 the interference between the diagrams is small and
the production cross section at LEP is $\sigma\sim 6-18~pb$
for $M_{{\tilde \chi}_1^+}\sim 60~GeV$.
Hence any limits which may be obtained will depend on
${\tilde m}_{\nu_L}$, as well as $\mu$ and $M_2$.   

The search proceeds by looking for the decay
${\tilde \chi}^\pm_1 \rightarrow {\tilde\chi}^0_1 
l^\pm \nu$
. 
The assumption is made that the ${\tilde \chi}^0_1$ is
stable and escapes the detector unseen.   Using this 
technique, ALEPH obtains a limit,\cite{aleph}
\beq
M_{\tilde \chi^\pm}> 67.8~GeV\quad @95\% CL
\nonumber 
\eeq
 
\beq
{\rm For}:\quad \quad \left\{ \begin{array}{l}
~m_{\tilde \nu_L} > 200~GeV
\\
 M_{\tilde\chi^\pm} ~{\rm gaugino-like}
, ~~\mid \mu\mid >> M_2   \quad .
  \end{array}
\right . 
\eeq
This limit is not very sensitive to $\tan\beta$,
but is considerably weaker when $\mid \mu\mid\le 100~GeV$.
It is clearly important to understand the  input assumptions about the
various SUSY parameters when interpreting this limit, as is
the case with most limits on SUSY particles.

It is interesting to compare the search for charginos and 
neutralinos at LEP with what is possible at the LHC.  At
the LHC one clear signature will be,\cite{charg} 
\beq
pp\rightarrow {\tilde \chi}^\pm_1 {\tilde \chi}^0_2
\nonumber 
\eeq
with,
\beqn 
{\tilde \chi}^\pm_1&\rightarrow &l^{\prime\pm} \nu {\tilde \chi}_1^0
\nonumber \\  
{\tilde \chi}^0_2&\rightarrow& l {\overline l} {\tilde \chi}_1^0
\quad . 
\eeqn
The cross section for this process is $\sigma \sim 1-100~pb$  
for masses  below  $1~TeV$.
This gives a ``tri-lepton signature" with three hard, isolated leptons,
significant $E_T$ and little jet activity.\cite{tri}
  The dominant Standard 
Model backgrounds are from $t {\overline t}$ production
(which  can be eliminated by requiring that the 2 fastest
leptons have the same sign) and $W^\pm Z$ production (which
is eliminated by requiring that $M_{ll}\ne M_Z$).

To get reliable predictions at a hadron collider, it is not 
enough to use your Monte Carlo generator
to simulate the process of interest (here chargino pair production).
One must also simulate all the other
 SUSY production processes.\cite{multi} 
It is amusing to note that at the LHC the largest background
to chargino and neutralino production is indeed   from other
SUSY particles, such as squark and gluino production,
which also give events with leptons, multi-jets, and missing $E_T$.
  Since the
squarks and gluinos are strongly interacting, they will generate
more jets and a harder missing $E_T$ spectrum than the charginos
and neutralinos.
This    
 can  be used to separate 
squark and gluino production from the chargino and neutralino
production process of interest.\cite{fp} 
\begin{itemize}
\item The biggest background to SUSY is SUSY itself.
\end{itemize}

As an example, we quote from a study
of the tri-lepton signature at the LHC
 which assumes
relatively light charginos and neutralinos,\cite{fp} 
\beqn
M_{{\tilde \chi}_1^+}&=& 96~GeV 
\nonumber \\
M_{{\tilde \chi}_2^0}&=& 96~GeV
\nonumber \\ 
M_{{\tilde \chi}_1^0}&=& 45~GeV
\quad .
\eeqn  
Once  the SUSY particle masses are specified
 all the production rates can be computed unambiguously.
After cuts, Ref. \cite{fp} finds (at the LHC):
\beqn
{\rm Signal}:  &&41~fb
\nonumber \\
t {\overline {t}}~ {\rm bkdg}:& & 2.4~ fb
\nonumber \\
WZ~ {\rm bkgd}:&& .5 fb
\nonumber \\
{\tilde g}, {\tilde q}~ {\rm bkgd}: &  & 5.6~fb
,  
\eeqn
demonstrating the viability of this signature at the LHC.

\begin{figure}[htb]
\vspace*{1.in}
\epsfig{file=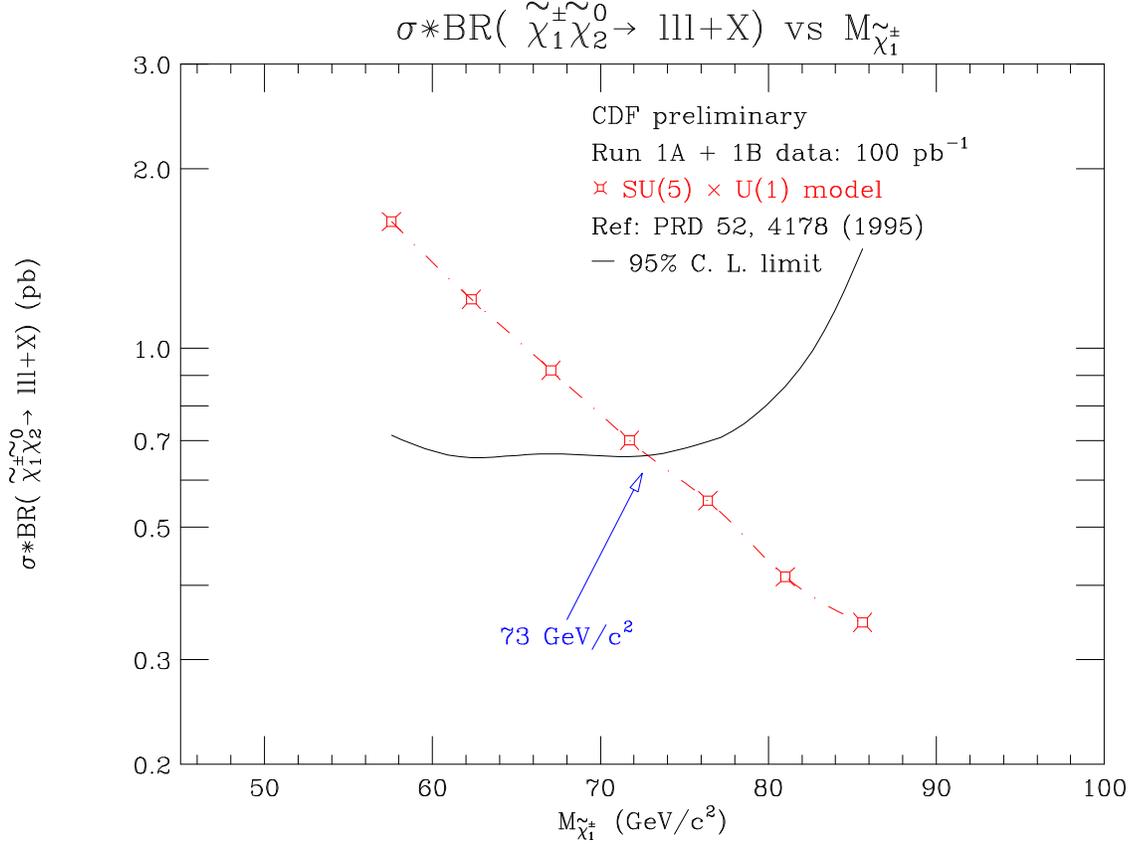,height=4.5in}
\caption{Limits on the tri-lepton
signature in the reaction $p {\overline p}\rightarrow
{\tilde \chi}^\pm_1{\tilde \chi}^0_2$ from CDF.  
This figure from Ref. [67].}
\vspace*{.5in}  
\end{figure}  
  
\begin{itemize}
\item
The tri-lepton signal offers the possibility of untangling the
${\tilde \chi}^+{\tilde \chi}^0$ signal from the gluino
and squark background.  
\end{itemize}  
CDF has searched for this decay chain and we see the results
in Fig. 16.\cite{cdflim}  Since the branching ratio to tri-leptons
depends on the parameters of the chargino and
neutralino mass matrix they also show the prediction from
a specific Grand Unified Theory.  Within this model, the
limit translates to $M_{{\tilde \chi}_1^\pm}> 73~GeV$.
This is roughly the same limit as that found at LEP, but
involves different assumptions about the parameters of the
model.     

Aside from observing the process and verifying
the existence of charginos and neutralinos,
 we would also like to obtain a 
handle on the masses of the SUSY particles.
  The kinematics are such that,
\beq
0 < M_{ll} < M_{{\tilde \chi}^0_2}-M_{{\tilde \chi}^0_1}
\quad ,
\eeq
and hence the distribution $d\sigma/dM_{ll}$ has a sharp cut-off
at the kinematic boundary which can be used to
obtain information on the masses.  Recently, significant
progress has been made in our understanding of the capabilities
of a hadron collider for extracting values of the SUSY particle
masses from different event distributions.\cite{fpsnow} 

\subsection{Squarks, Gluinos, and Sleptons}
 
Squarks and sleptons,(${\tilde f}_i$),  can be
 produced at both $e^+e^-$ and
hadron colliders.  At LEP, they would be  pair produced via
\beq
e^+e^-\rightarrow \gamma,Z
\rightarrow {\tilde f}_i {\tilde f}^*_i.
\eeq 
If there were a scalar with mass  less than half the
$Z$ mass,  it would 
increase the total width of the $Z$, $\Gamma_Z$.
Since $\Gamma_Z$ agrees quite precisely with the
Standard Model prediction,  
the measurement of the $Z$ lineshape gives
\beq
{\tilde m}> 35-40~GeV\eeq  
 for squarks and sleptons.  
The limit from the $Z$ width is  
particularly important  because it is independent
of the squark or slepton decay mode and so applies for
any model with low energy supersymmetry.

\begin{figure}[htb] 
\centerline{\epsfig{file=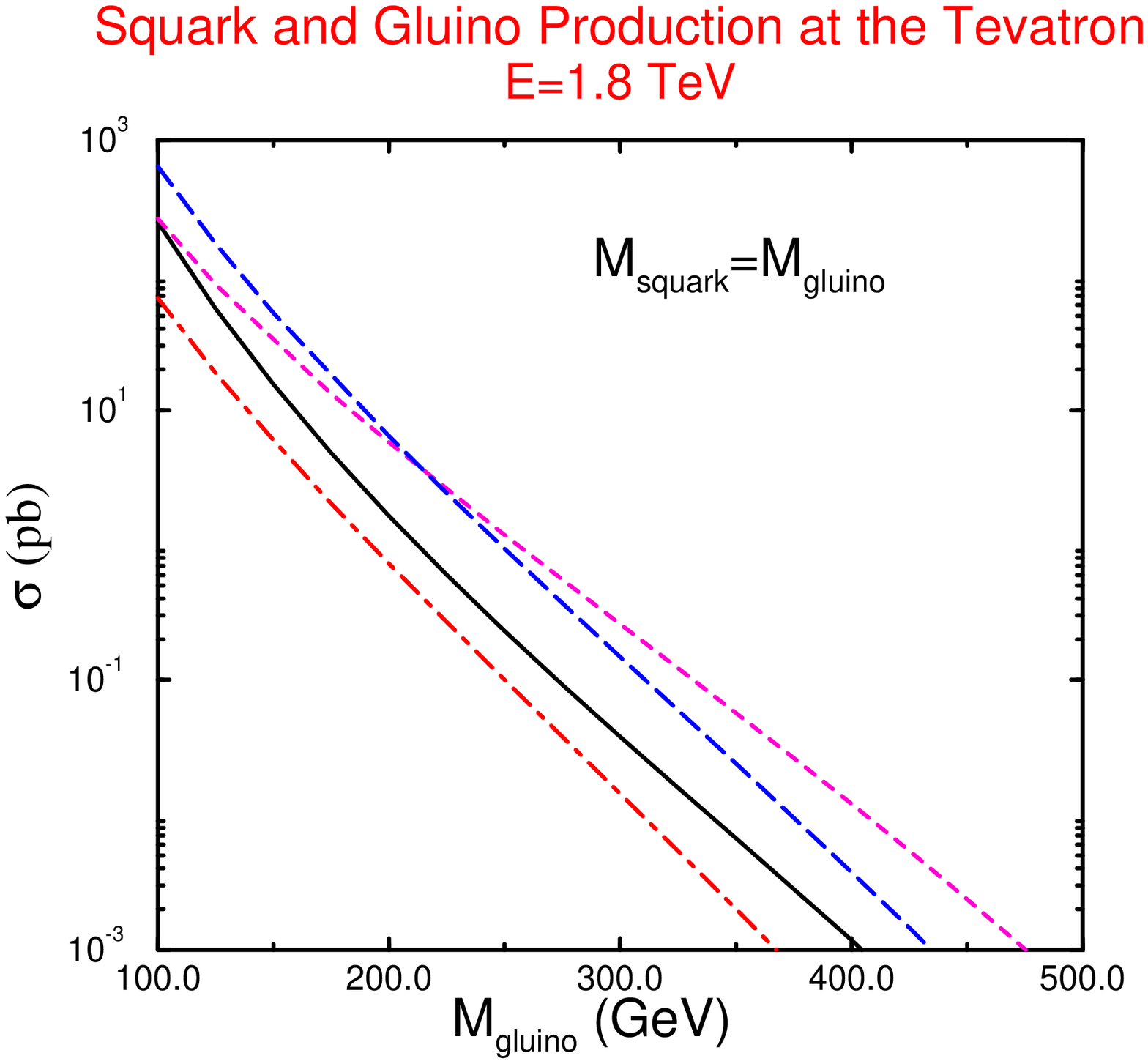,height=4.in}}
\caption{Squark and gluino production
at the Tevatron assuming $M_{\tilde q}=M_{\tilde g}$.
  The solid line is $p {\overline p}
\rightarrow {\tilde g}{\tilde g}$, the dot-dashed
${\tilde q}{\tilde q}$,
 the dotted ${\tilde q}{\tilde q}^*$,
 and the dashed is ${\tilde q}{\tilde g}$. This figure includes only
the Born result and assumes $10$ degenerate squarks.}
\end{figure}
  
\begin{figure}[htb]
\vspace*{1.in} 
\centerline{\epsfig{file=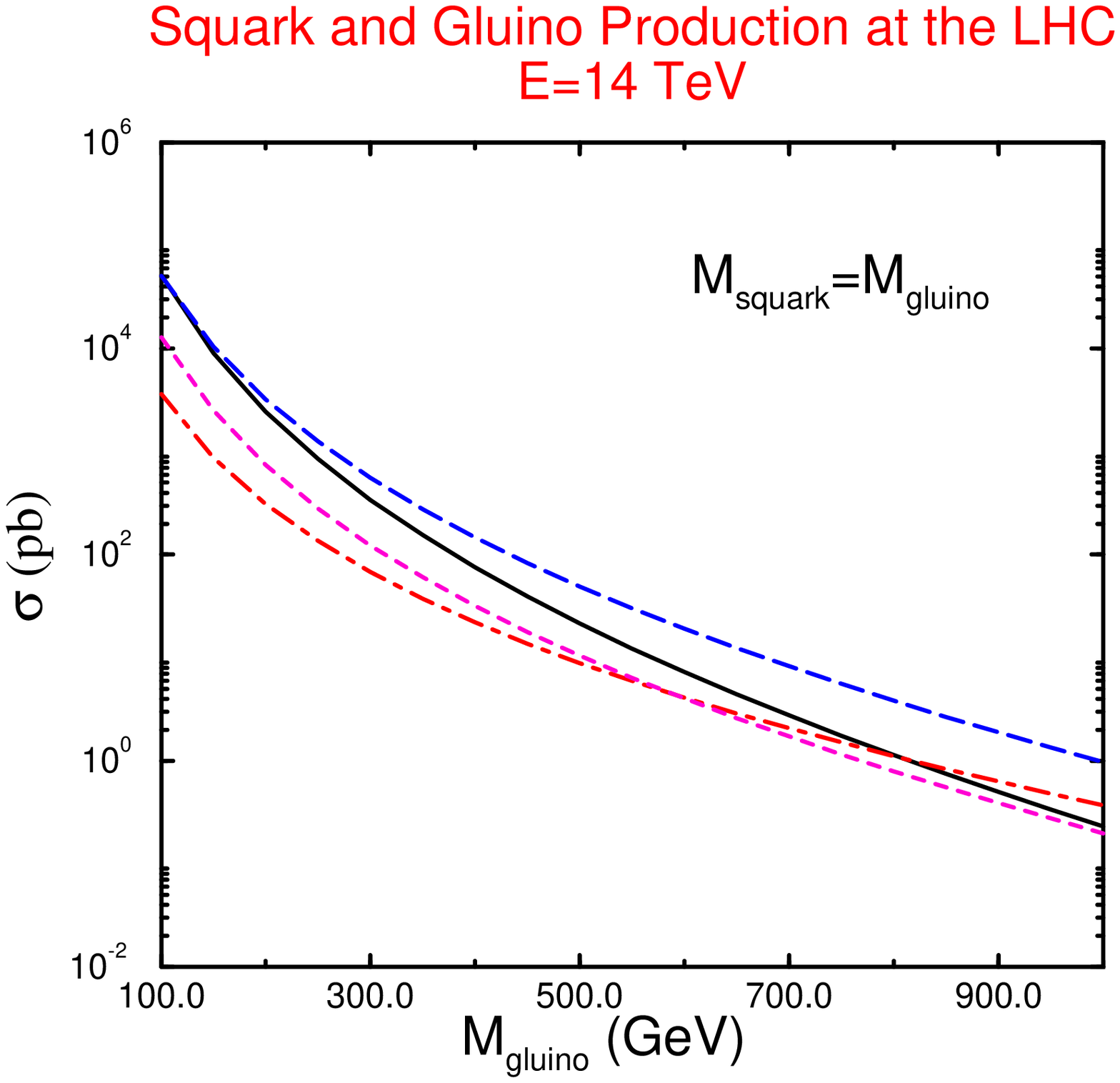,height=4.in}}
\caption{Squark and gluino production at the LHC 
assuming $M_{\tilde q}=M_{\tilde g}$.
The solid line is $ p {\overline p}
\rightarrow {\tilde g}{\tilde g}$, the dot-dashed 
${\tilde q}{\tilde q}$, the dotted ${\tilde q}{\tilde q}^*$,
and the dashed is ${\tilde q}{\tilde g}$.  This figure
includes only the Born result and assumes $10$ degenerate 
squarks.}
\vspace*{.5in} 
\end{figure}

There are limits on the direct production of
 squarks and gluinos from
the Tevatron.  The rates for squark and
gluino production at both the Tevatron and the
  LHC are shown in Figs. 17 and 18 and analytic expressions for
the Born cross sections can be found in Ref. \cite{sigsusy}.
The QCD radiative corrections to these process are
large and increase the cross sections 
by up to a factor of two.\cite{squglu} 
 We neglect the mixing
effects in the squark mass matrix and assume that there are $10$
degenerate squarks associated with the light quarks.  
(The top squarks are assumed to be different since here 
mixing effects are clearly relevant.) 
   The cross sections are significant, around
$1~pb$ for squarks and gluinos in the few hundred GeV range.

The cleanest signatures for squark and gluino production
are jets plus missing $E_T$ from
the undetected  LSP, 
assumed to be ${\tilde \chi}_1^0$,
 and jets plus multi leptons
plus missing $E_T$.\cite{squark}
  It will clearly be exceedingly difficult
to separate the effects of squarks and gluino production,
since they both contribute to the same experimental signature.
The patterns of squark decays in various scenarios are
examined in Ref. \cite{squarkdecay}.    
To obtain a limit on the gluino mass, we must 
therefore assume a limit
on the squark mass.    
 For 10 degenerate squarks, the limit from the Tevatron is, \cite{mer,pdg} 
\beq
M_{\tilde q}> 218~GeV\qquad {\rm for} M_{\tilde g}=M_{\tilde q}
\quad .
\eeq
This limit assumes a cascade decay, ${\tilde q}\rightarrow
(....) {\tilde \chi}_1^0$. (There are similar limits for
$M_{\tilde g}<<M_{\tilde q}$
 and $M_{\tilde g}>>M_{\tilde q}$.)

Limits on the stop squark are particularly interesting since in many models 
it is the lightest squark.  There are 2 types of stop squark decays which 
are relevant.  The first is,
\beq
{\tilde t}\rightarrow b {\tilde \chi}^+_1 
\rightarrow b  f {\overline f}^\prime {\tilde \chi}_i^0
\quad .
\eeq 
The signal for this decay channel  is jets plus missing 
energy.  This signal shares many features with the
dominant
top quark decay, $t\rightarrow b W^+$, and in fact there have
been  suggestions in the literature that there may be
some experimental 
confusion between the $2$ processes.\cite{barn}  Another possible decay
chain for the stop squark
is 
\beq
{\tilde t}\rightarrow c {{\tilde \chi}_1^0},
\eeq
which also leads to jets plus missing energy.  The $2$ cases
must be analyzed separately.  The current limit
on the stop squark mass  from D0 is shown in Fig. 19.\cite{d0lim} 
We see that the limit depends sensitively on the mass of the
LSP, ${\tilde \chi}_1^0$.
\begin{figure}[htb]
\epsfig{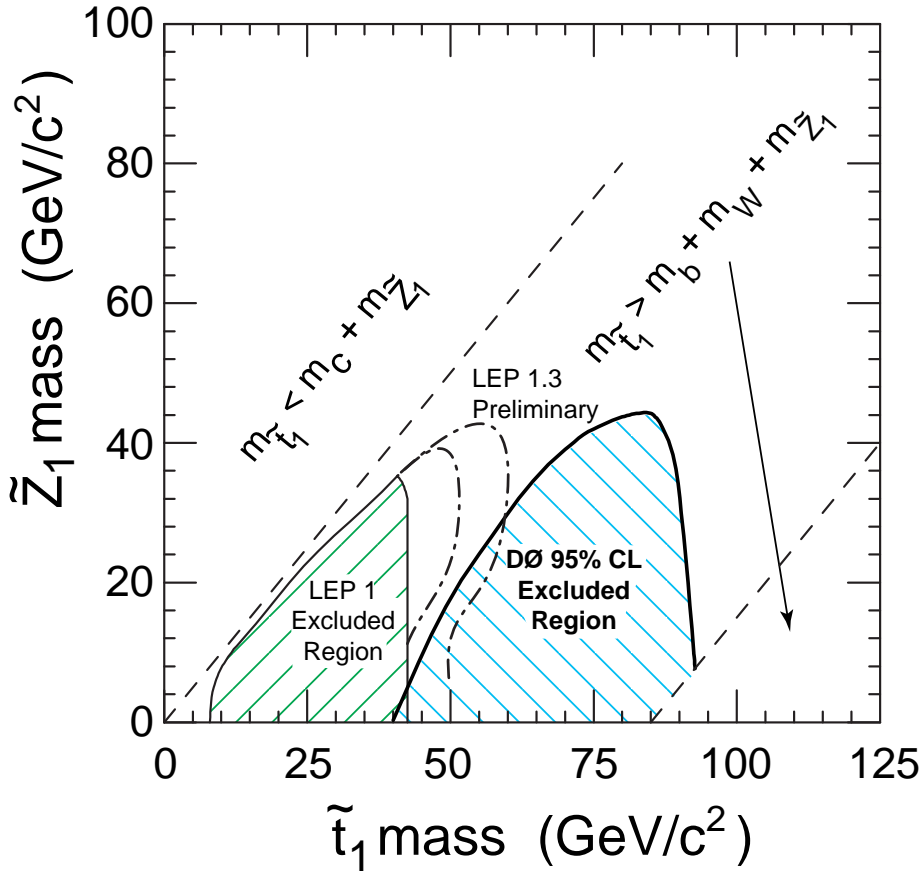}
\caption{Limit on the lightest stop squark mass
here labelled ${\tilde t}_1$, as a function of
the lightest chargino mass (here labelled ${\tilde Z}_1$)
from D0.  This figure from Ref. [74].}  
\end{figure}

A spectacular signal for squark pair production which can
result from the cascade decays is the production of same sign
leptons,
\beq
p p\rightarrow {\tilde q}{\tilde q}^*
\rightarrow (l^\pm l^\pm)+{\rm jets} + E_T^{\rm miss}
\eeq
At the Tevatron with $M_{\tilde q}\sim M_{\tilde g}\sim 100~GeV$,
the cross section for jets + $E_T^{\rm miss}$ is $\sigma \sim 1~pb$,
while the rate for $l^\pm l^\pm + {\rm ~jets} + E_T^{\rm miss}$
is $\sigma \sim .1 pb$, which is still significant.  The Standard
Model background for this signal is quite small.    

 From the examples we have given, it is clear that searching for
SUSY at a hadron collider is  particularly
challenging since there will
typically be many SUSY particles which are kinematically
accessible.  Hadron colliders  thus have a large discovery
potential, but it is difficult to separate the various processes.
To a large extent, one must trust the generic signatures of
supersymmetry:  $E_T^{\rm miss}$, plus multi-jet and multi-lepton
signatures.  One will need to observe a signal in many channels in
order to verify the consistency of the model.

\subsection{A Case Study}

It is instructive  to consider an example of how
the discovery of a SUSY particle might occur.  Several
years ago, CDF presented a single event,
\beq
p {\overline p} \rightarrow e^+e^-\gamma\gamma +
 E_T^{miss}
,
\eeq
for which it was difficult to find a Standard Model
explanation.\cite{cdfdata}  By now, you all know
that events with large missing energy are candidates
for SUSY particle production.  The scenario which
we  can construct is then,
\beq
p {\overline p} \rightarrow {\tilde e}{\tilde e}^*
\quad ,
\eeq
where ${\tilde e}$ is the scalar partner of either the
right- or left-handed electron.  The production cross section
is then fixed unambiguously in terms of the selectron mass.
The fact that only
one event was seen fixes the selectron mass to be in the
$100~GeV$ region.  
The selectron is then assumed to decay to an electron and
a neutralino,
\beq
{\tilde e}\rightarrow e {\tilde \chi}^0
\quad .
\eeq

The question which  has engendered furious debate is how the
 neutralino   might decay,
\beq
{\tilde \chi}^0 \rightarrow {\tilde X} \gamma,
\eeq
where ${\tilde X}$ is either  the lightest  neutralino or a 
gravitino.\cite{cdfevent}  
By examining the kinematics of the event, we could hope to
learn about the underlying SUSY model.  Unfortunately, examination
of the $2$ photon plus $E_T^{miss}$ spectrum has produced no more SUSY 
candidates of the type of Eq. 94.\cite{cdfgam} 

\section{CONCLUSIONS}  

Weak scale supersymmetry is a theory in
desperate need of experimental input.  The 
theoretical framework has evolved to a point where
predictions for cross sections, branching ratios, and decay
signatures can be reliably made.  In many cases,
calculations exist beyond the leading order in perturbation
theory.   However, without
experimental observation of a SUSY particle or a
precision measurement which disagrees with the Standard
Model (which could be explained by
SUSY particles in loops) there is no way of
choosing between the many possible manifestations of low
energy SUSY and thereby  fixing the parameters in the soft SUSY
breaking Lagrangian. 

With the coming of LEPII, the Fermilab Main Injector, and the
LHC, large regions of SUSY parameter space will be 
explored and we can only hope that some evidence for
supersymmetry will be uncovered.  
 The ball is definitely in the
experimentalist's court~!

\subsection{Acknowledgments}

I thank all the students at this school who asked such
wonderful questions and really made me think about 
the experimental consequences of supersymmetry
I also thank
Tom Ferbel for his superb organization of all the lectures
and other happenings at this school.
I am grateful to Michael Spira for the use of his program
HDECAY for the calculation of the radiatively
corrected SUSY Higgs decay widths 
and also for discussions.  
Helpful discussions with Frank Paige are also  gratefully
acknowledged.
I am indebted to Ken Kiers for a careful reading of the
manuscript.  
This work has been supported by the DOE under contract
number DE-AC02-76-CH-00016.

\end{document}